\pdfoutput=1
\documentclass[twocolumn,english,aps,amsmath,amssymb,superscriptaddress]{revtex4-1}

\usepackage{babel}
\usepackage{amsmath}
\usepackage{amssymb}
\usepackage{graphicx}
\usepackage{xcolor}
\usepackage{tikz}
\usepackage[utf8]{inputenc}

\begin{document}

\title{Identifying the Huse-Fisher universality class of the three-state chiral Potts model}

\author{Samuel Nyckees}
\affiliation{Institute of Physics, Ecole Polytechnique F\'ed\'erale de Lausanne (EPFL), CH-1015 Lausanne, Switzerland}
\author{Jeanne Colbois}
\affiliation{Institute of Physics, Ecole Polytechnique F\'ed\'erale de Lausanne (EPFL), CH-1015 Lausanne, Switzerland}
\author{Fr\'ed\'eric Mila}
\affiliation{Institute of Physics, Ecole Polytechnique F\'ed\'erale de Lausanne (EPFL), CH-1015 Lausanne, Switzerland}

\date{\today}
\begin{abstract} 
Using the corner-transfer matrix renormalization group approach, we revisit the three-state chiral Potts model on the square lattice, a model proposed in the eighties to describe commensurate-incommensurate transitions at surfaces, and with direct relevance to recent experiments on chains of Rydberg atoms.  This model was suggested by Huse and Fisher to have a chiral transition in the vicinity of the Potts point, a possibility that turned out to be very difficult to definitely establish or refute numerically. Our results confirm that the transition changes character at a Lifshitz point that separates a line of Pokrosky-Talapov transition far enough from the Potts point from a line of direct continuous order-disorder transition close to it. Thanks to the accuracy of the numerical results, we have been able to base the analysis entirely on effective exponents to deal with the crossovers that have hampered previous numerical investigations. The emerging picture is that of a new universality class with exponents that do not change between the Potts point and the Lifshitz point, and that are consistent with those of a self-dual version of the model, namely correlation lengths exponents $\nu_x=2/3$ in the direction of the asymmetry and $\nu_y=1$ perpendicular to it, an incommensurability exponent $\bar \beta=2/3$, a specific heat exponent that keeps the value $\alpha=1/3$ of the three-state Potts model, and a dynamical exponent $z=3/2$. These results are in excellent agreement with experimental results obtained on reconstructed surfaces in the nineties, and shed light on recent Kibble-Zurek experiments on the period-3 phase of chains of Rydberg atoms.
\end{abstract}

\maketitle


\section{Introduction}

Since its introduction by Ostlund\cite{ostlund} and Huse\cite{huse} in the context of commensurate-incommensurate transitions, the chiral Potts model has been the focus of an uninterrupted activity both in its two-dimensional statistical physics formulation\cite{ostlund,huse,HuseFisher, Selke1982,haldane_bak,schulz,Duxbury,nijs1984,HuseFisher1984,yeomans1985,houlrik1986,bartelt,stella1987,auyang1987,baxter1988,Everts_1989,Baxter1989,albertini,mccoy,cardy,auyang1996,sato_sasaki,fendley_parafermions,sachdev_dual}, and in its one-dimensional quantum version\cite{CENTEN1982585,HOWES1983169,howes1983,vongehlen,fendley,sachdev_girvin,Lesanovsky2012,hughes,lukin2017,samajdar,lukin2019,chepiga_mila_PRL,dalmonte2019}. When the asymmetry parameters are allowed to take arbitrary complex values, this defines a family of models, some of them with complex Botzmann weights, and several exact results have been obtained over the years\cite{Baxter1989,albertini,mccoy,cardy}. In particular, there is a two-parameter family of integrable models with rather unusual properties\cite{Baxter1989,albertini,mccoy}. 

However, the physical properties of the chiral 3-state Potts model introduced by  Ostlund and Huse are not fully understood. This model is defined in terms of local variables $n_{\vec r}=0,1,2$ on a square lattice by the energy
\begin{eqnarray}
E=-\sum_{\vec r} \cos[2\pi/3(n_{\vec r + \vec x}-n_{\vec r}+\Delta)]\nonumber\\
-\sum_{\vec r} \cos[2\pi/3(n_{\vec r + \vec y}-n_{\vec r})] 
\label{eq:model}
\end{eqnarray}
where $\vec x$ and $\vec y$ are the basis vectors of the lattice. For this model, with only a real asymmetry parameter $\Delta$ in one direction, there is no exact solution except at the Potts point $\Delta=0$. At that point, the critical temperature is known exactly from a duality argument, $T_c=3/[2\ln(\sqrt{3}+1)]$, and the correlation length diverges with an exponent $\nu=5/6$. Away from this point, the chiral perturbation introduced by $\Delta$ is relevant, and the transition has to be modified in an essential way. One possibility is that a critical floating phase opens immediately, bounded by a Kosterlitz-Thouless transition\cite{Kosterlitz_Thouless} at high temperature and a Pokrovsky-Talapov transition\cite{Pokrovsky_Talapov,schulz1980} at low temperature. However, Huse and Fisher suggested in 1982 that the transition could remain a direct commensurate-incommensurate transition up to a Lifschitz point L, but in a new chiral universality class characterized by $q \,\xi_x \rightarrow C>0$, where $\xi_x$ is the correlation length in the $x$ direction, and $q$ is the incommensurate vector in the high temperature phase\cite{HuseFisher}. More precisely, if $\xi_x$  diverges as $1/t^{\nu_x}$, where $t=(T-T_c)/T_c$, and $q$ vanishes as $t^{\bar \beta}$, then this universality class would be characterized by $\nu_x=\bar \beta$, by contrast to the Potts point, where $\nu_x=5/6$  and  $\bar \beta=5/3$. While all numerical results seem to be consistent with a single transition for not too large $\Delta$, it has proven exceedingly difficult to determine these critical exponents, either numerically with Monte Carlo\cite{Selke1982,houlrik1986,bartelt}, or using finite-size renormalization group\cite{Duxbury,yeomans1985} or finite-size transfer matrix\cite{stella1987,Everts_1989}, and the question remains unsettled as to whether there is indeed a chiral transition, or rather a very narrow floating phase up to the Potts point.

\begin{figure}[t!]
\includegraphics[width=0.45\textwidth]{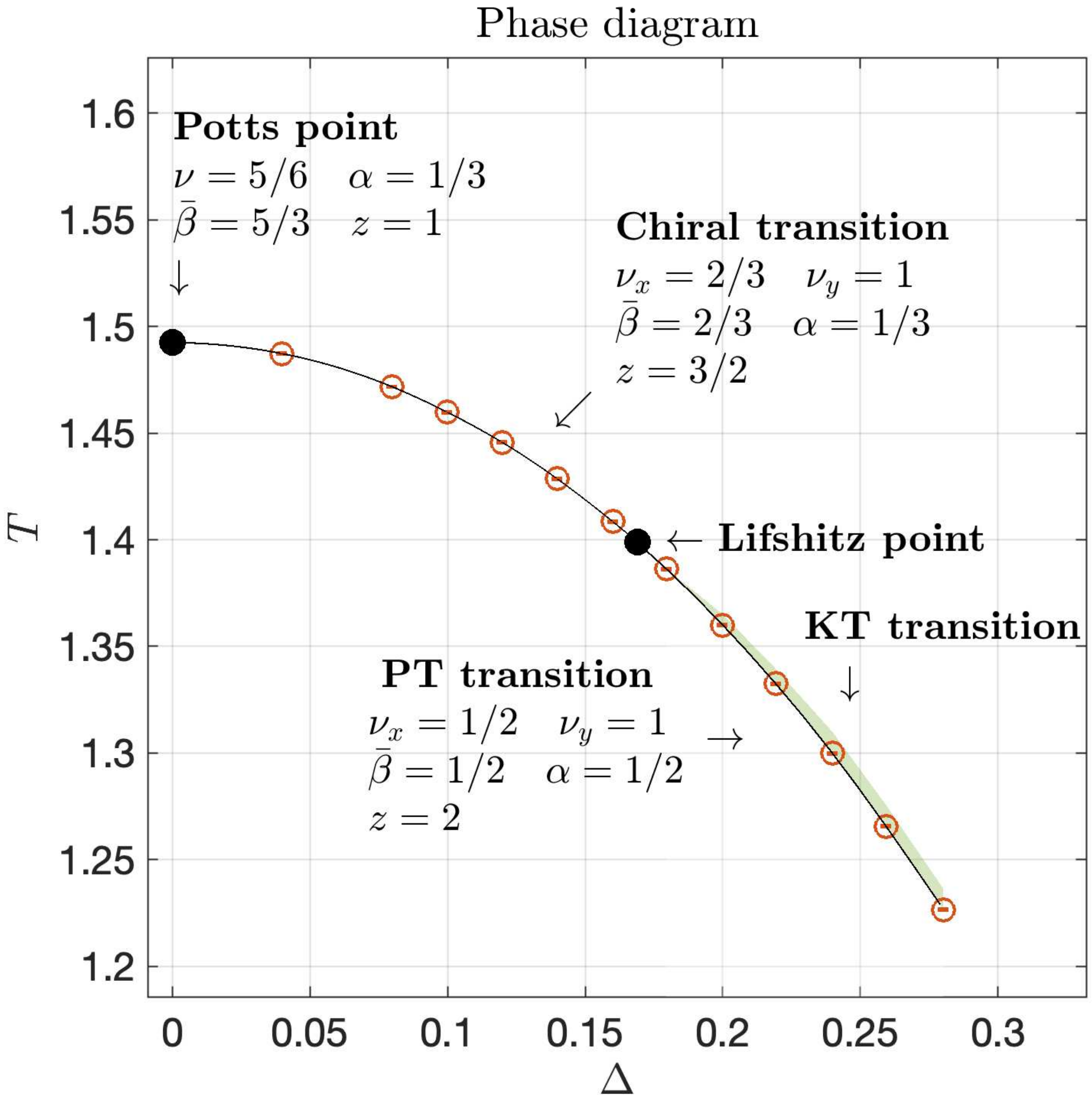}
\caption{Phase diagram of the chiral three-state Potts model of Eq. \ref{eq:model}. The precision on the critical temperature is about $10^{-4}$. The Lifshitz point is located at $\Delta=0.169\pm0.003$. The values of the critical exponents are the known exact values at the Potts point and for the Pokrovsky-Talapov transition, and our conjectured values based on our numerical analysis for the Huse-Fisher chiral transition. The extent of the critical region shown in green is purely indicative.
}
\label{fig:phase_diagram}
\end{figure}

In this paper, we revisit this issue using a more recent numerical approach, the corner transfer matrix renormalization group (CTMRG), and 
a different strategy. Rather than trying to measure the product $q\,\xi_x$ along the transition line, which, as we shall see, is not precise enough because of severe and incompatible crossover regimes, we study separately the scaling of the correlation lengths $\xi_x$ and $\xi_y$, of the wave vector $q$, and of the specific heat $C$ assuming that crossover regimes have to be overcome to reveal the true critical behaviour of these quantities. To achieve this, we systematically study the behaviour of effective exponents close to the transition, a method introduced in the study of imperfect surfaces\cite{pleimling_selke} and proven to be useful in the study of the Lifshitz point of the ANNNI model\cite{pleimling_henkel}. As we shall see, the picture that emerges is that of a unique universality class between the Potts point and a Lifshitz point at $\Delta\simeq 0.17$ characterized by exponents that are consistent with those of the integrable version of the model $\nu_x=2/3$, $\nu_y=1$, $\bar \beta=2/3$, and $\alpha=1/3$. This suggests that, as anticipated by Huse and Fisher, the transition away from the Potts point and up to a Lifshitz point is in a non-conformal universality class with a dynamical exponent $z=3/2$. Beyond the Lifshitz point, we confirm that the commensurate-incommensurate transition is in the Pokrovsky-Talapov universality class, and that the correlation length diverges before that transition when coming from the disorder side, consistent with a Kosterlitz-Thouless transition into a floating phase. The phase diagram is summarized in Fig. \ref{fig:phase_diagram}.

The paper is organized as follows. In Section II, we review the CTMRG method, and we explain our strategy based on effective exponents and on a very careful estimate of the critical temperature constrained by physical considerations. In Section III, we present the main numerical results obtained in this paper, first far away from the Potts point, then in its vicinity, with evidence of a Lifshitz point on the way. These results are critically reviewed in Section III, with an analysis based on scaling relations that point to a unique chiral universality class with strong crossover effects, and with a comparison with experimental results on reconstructed surfaces and on chains of Rydberg atoms. The results are summarized and put in perspective in Section IV.

\section{Methodology}

The analysis carried out in this paper relies heavily on three main ingredients: i) the formulation of the partition function as a tensor network and its approximate contraction using CTMRG; ii) the analysis of effective exponents defined as the local slope of the log-log plot of a quantity as a function of the reduced temperature $t$; iii) a strategy to estimate the critical temperature based on the possible nature of the phase transitions. 

\subsection{CTMRG}

The starting point is the observation that one can write the partition function of the asymmetric three-state Potts model as a contraction of an infinite two dimensional square tensor network, with, laying on each vertex, a $3\times3\times3\times3$ dimensional tensor $a$,
\begin{align*}
   a_{ijkl}= \sum_{\alpha} (\sqrt{Q^x})_{i\alpha} (\sqrt{Q^y})_{j \alpha}  (\sqrt{Q^x})_{\alpha k} (\sqrt{Q^y})_{\alpha l} 
\end{align*}
with
\begin{align*}
Q^y = \begin{pmatrix}
  e^{\beta J \text{cos}(\frac{2\pi}{3}\Delta)} & e^{\beta J \text{cos}(\frac{2\pi}{3}(\Delta-1))} & e^{\beta J \text{cos}(\frac{2\pi}{3}(\Delta-2))}  \\
  e^{\beta J \text{cos}(\frac{2\pi}{3}(\Delta+1))} &  e^{\beta J \text{cos}(\frac{2\pi}{3}\Delta)} &  e^{\beta J \text{cos}(\frac{2\pi}{3}(\Delta-1))}\\
  e^{\beta J \text{cos}(\frac{2\pi}{3}(\Delta+2))} & e^{\beta J \text{cos}(\frac{2\pi}{3}(\Delta+1))} & e^{\beta J \text{cos}(\frac{2\pi}{3}\Delta)} 
 \end{pmatrix}
\end{align*}
and the same expression for $Q^x$ with $\Delta=0$.

The CTMRG algorithm that we will use systematically throughout was first introduced by Nishino and Okunishi\cite{nishino}. It is a combination of Baxter's corner transfer matrix\cite{baxter1968,baxter1978} and of the density matrix renormalization group algorithm\cite{dmrg1,white1993}. This algorithm approximates the partition function by reducing the infinite contraction problem to the contraction of nine tensors. The size of the tensors is controlled by a bond dimension parameter $\chi$, which, when taken to infinity gives an exact result. The symmetries of the model allow us to reduce the number of tensors to six: $\mathcal{T}=\{C_R, C_L, T^y_L, T^y_R, T^x, a\}$. Within this formalism, the computation of every local observable such as the energy is reduced to the contraction of relatively small networks. 

\begin{center}
\begin{tikzpicture}
\draw (-5, -1) node{$\mathcal{Z}= $};
\draw (-0.9, -1) node{$\approx$};
\draw (0.5, 0.2) node{$\chi$};
\draw (-3.6, -0.2) circle (0.2cm) node{$a$}; \draw (-2.8, -0.2) circle (0.2cm) node{$a$}; \draw (-2, -0.2) circle (0.2cm) node{$a$}; 
\draw (-3.6, -1) circle (0.2cm) node{$a$}; \draw (-2.8, -1) circle (0.2cm) node{$a$}; \draw (-2, -1) circle (0.2cm) node{$a$}; 
\draw (-3.6, -1.8) circle (0.2cm) node{$a$}; \draw (-2.8, -1.8) circle (0.2cm) node{$a$}; \draw (-2, -1.8) circle (0.2cm) node{$a$}; 
\draw  [dashed] (-4.3, -0.2)--(-3.8, -0.2);\draw (-3.4, -0.2) -- (-3, -0.2); \draw (-2.6, -0.2)--(-2.2,-0.2); \draw [dashed] (-1.8,-0.2) -- (-1.3,-0.2);
\draw  [dashed] (-4.3, -1)--(-3.8, -1);\draw (-3.4, -1) -- (-3, -1); \draw (-2.6, -1)--(-2.2,-1); \draw [dashed] (-1.8,-1) -- (-1.3,-1);
\draw  [dashed] (-4.3, -1.8)--(-3.8, -1.8);\draw (-3.4, -1.8) -- (-3, -1.8); \draw (-2.6, -1.8)--(-2.2,-1.8); \draw [dashed] (-1.8,-1.8) -- (-1.3,-1.8);
\draw [dashed] (-3.6, 0)--(-3.6, 0.5); \draw [dashed](-2.8, 0)-- (-2.8, 0.5); \draw [dashed](-2, 0)-- (-2, 0.5);
\draw (-3.6, -0.4)-- (-3.6, -0.8); \draw (-2.8, -0.4)-- (-2.8, -0.8); \draw (-2, -0.4)-- (-2, -0.8);
\draw (-3.6, -1.2)-- (-3.6, -1.6); \draw (-2.8, -1.2)-- (-2.8, -1.6); \draw (-2, -1.2)-- (-2, -1.6);
\draw [dashed] (-3.6, -2)--(-3.6, -2.5); \draw [dashed](-2.8, -2)-- (-2.8, -2.5); \draw [dashed](-2, -2)-- (-2, -2.5);
\draw (0,0) circle (0.3cm) node{$C_L$}; \draw (1,0) circle (0.3cm) node{$T^x$}; \draw (2,0) circle (0.3cm) node{$C_R$};
\draw (0,-1) circle (0.3cm) node{$T^y_L$}; \draw (1, -1) circle(0.2cm) node{$a$}; \draw (2,-1) circle (0.3cm) node{$T^y_R$};
\draw (0,-2) circle (0.3cm) node{$C_L$}; \draw (1,-2) circle (0.3cm) node{$T^x$}; \draw (2,-2) circle (0.3cm) node{$C_R$};

\draw [line width=0.3mm] (0.3,0) --  (0.7,0); \draw [line width=0.3mm] (1.3,0) -- (1.7,0);
\draw [line width=0.3mm] (0,-0.3)--(0,-0.7); \draw (1,-0.3) -- (1,-0.8); \draw  [line width=0.3mm] (2,-0.3) -- (2,-0.7);
\draw (0.3,-1) -- (0.8,-1); \draw (1.7,-1) -- (1.2,-1);
\draw [line width=0.3mm] (0,-1.3) -- (0,-1.7);  \draw (1,-1.7) -- (1, -1.2);\draw [line width=0.3mm] (2,-1.3) -- (2,-1.7);
\draw  [line width=0.3mm] (1.7,-2) -- (1.3,-2); \draw   [line width=0.3mm](0.7,-2) -- (0.3,-2);
\end{tikzpicture}
\end{center}

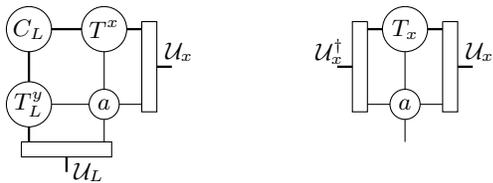
\begin{figure}
\begin{center}
\begin{tikzpicture}
\draw (0,0) circle (0.3cm) node{$C_L$}; \draw (1,0) circle (0.3cm) node{$T^x$};
\draw (0,-1) circle (0.3cm) node{$T^y_L$}; \draw (1,-1) circle (0.2cm) node{$a$};
\draw [line width=0.3mm] (0.3, 0) -- (0.7,0); \draw  [line width=0.3mm](1.3,0)--(1.5,0);
\draw [line width=0.3mm] (0,-0.3) -- (0,-0.7); \draw (1,-0.3)--(1, -0.8);
\draw  (0.3, -1) -- (0.8,-1); \draw (1.2,-1)--(1.5,-1);
\draw [line width=0.3mm] (0,-1.3)-- (0,-1.5); \draw (1,-1.2)--(1,-1.5);
\draw (1.5,0.1) -- (1.5,-1.1) -- (1.7,-1.1) -- (1.7,0.1) -- (1.5,0.1);
\draw [line width=0.3mm] (1.7, -0.5)--(1.9, -0.5);
\draw (2, -0.3) node{$\mathcal{U}_x$};
\draw (-0.1,-1.5) -- (-0.1,-1.7) -- (1.1,-1.7) -- (1.1,-1.5) -- (-0.1,-1.5);
\draw [line width=0.3mm] (0.5, -1.7)--(0.5, -1.9);
\draw (0.8, -1.9) node{$\mathcal{U}_L$};

\draw [line width=0.3mm] (4.5,0)--(4.7,0);  \draw (5,0) circle (0.3cm) node{$T_x$}; \draw [line width=0.3mm] (5.3,0)--(5.5,0); 
\draw (5,-0.3)--(5,-0.8);
\draw (4.5,-1)--(4.8,-1);  \draw (5,-1) circle (0.2cm) node{$a$}; \draw (5.2,-1)--(5.5,-1); 
\draw (5,-1.2)--(5,-1.5);
\draw (5.5, 0.1) -- (5.5,-1.1) -- (5.7, -1.1) -- (5.7,0.1) -- (5.5, 0.1);
\draw [line width=0.3mm] (5.7,-0.5)--(5.9,-0.5);
\draw (6, -0.3) node{$\mathcal{U}_x$};
\draw (4.5, 0.1) -- (4.3,0.1) -- (4.3, -1.1) -- (4.5,-1.1) -- (4.5, 0.1);
\draw [line width=0.3mm] (4.1,-0.5)--(4.3,-0.5);
\draw (4., -0.3) node{$\mathcal{U}_x^\dagger$};
\end{tikzpicture}
\caption{Full iteration for the corner and row tensor $C_L$ and $T^x$.\label{fig:Fig2}
}
\end{center}
\end{figure}

$\mathcal{T}$ is obtained through a two-step iterative process (Fig. \ref{fig:Fig2}) that goes on until the energy has converged to some precision, after which we consider the thermodynamic limit to be reached. 

(1) \textit{Extension:} the row tensors $T^x,T^y_L$ and $a$ are contracted with $C_L$ to form $\Tilde{C_L}$, and the row tensors $T^x,T^y_R$ and $a$ with $C_R$ to form $\Tilde{C_R}$. Similarly, the local tensor $a$ is contracted with each row tensor. The bond dimension of the row and corner tensors has increased by a factor 3. So, without some approximation, the dimension increases exponentially with the number of iterations.

(2) \textit{Truncation:} in order to reduce the dimension, each tensor has to be truncated. This truncation is done by unitary matrices $\mathcal{U}_i$, called isometries, that reduce the dimension of the tensors to $\chi$. Multiple choices of unitary matrices have been proposed in the literature. We use those suggested by Orus and Vidal\cite{orus}. They are constructed by applying the singular value decomposition on reduced density matrices defined with the corner transfer matrices $\Tilde{C_L}$ and $\Tilde{C_R}$. The symmetries allow us to use three different isometries.
\begin{align}
    &{\mathcal{U}_x}' S_x {{\mathcal{V}_x'}^\dagger} = \Tilde{C_L}^\dagger \Tilde{C_L} + \Tilde{C_R} \Tilde{C_R}^\dagger \nonumber \\
    &{\mathcal{U}_L}' S_L {{\mathcal{V}_L'}^{\dagger}} = \Tilde{C_L} \Tilde{C_L}^\dagger \nonumber \\
    &{\mathcal{U}_R}' S_R {{\mathcal{V}_R'}^{\dagger}} = \Tilde{C_R}^\dagger \Tilde{C_R} 
    \label{eq:isometries}
\end{align}
We then truncate the singular matrices $\mathcal{U}'_i$ into $\mathcal{U}_i$ by keeping only the $\chi$ largest singular values.

One of the main advantages of the method is that it gives direct access to the transfer matrices and their spectrum in both the $x$ and $y$ directions. We denote their normalized ($\lambda_1=1$) ordered eigenvalues as
\begin{align}
    \lambda_j = e^{-\epsilon_j-i \phi_j}, \qquad j\in \mathbb{N^*}
\end{align}
With this notation, the correlation length and the wave vector are given by
\begin{align}
   & \xi= \frac{1}{\epsilon_2}, \qquad q= \phi_2
\end{align}

\begin{figure}[t!]
\includegraphics[width=0.49\textwidth]{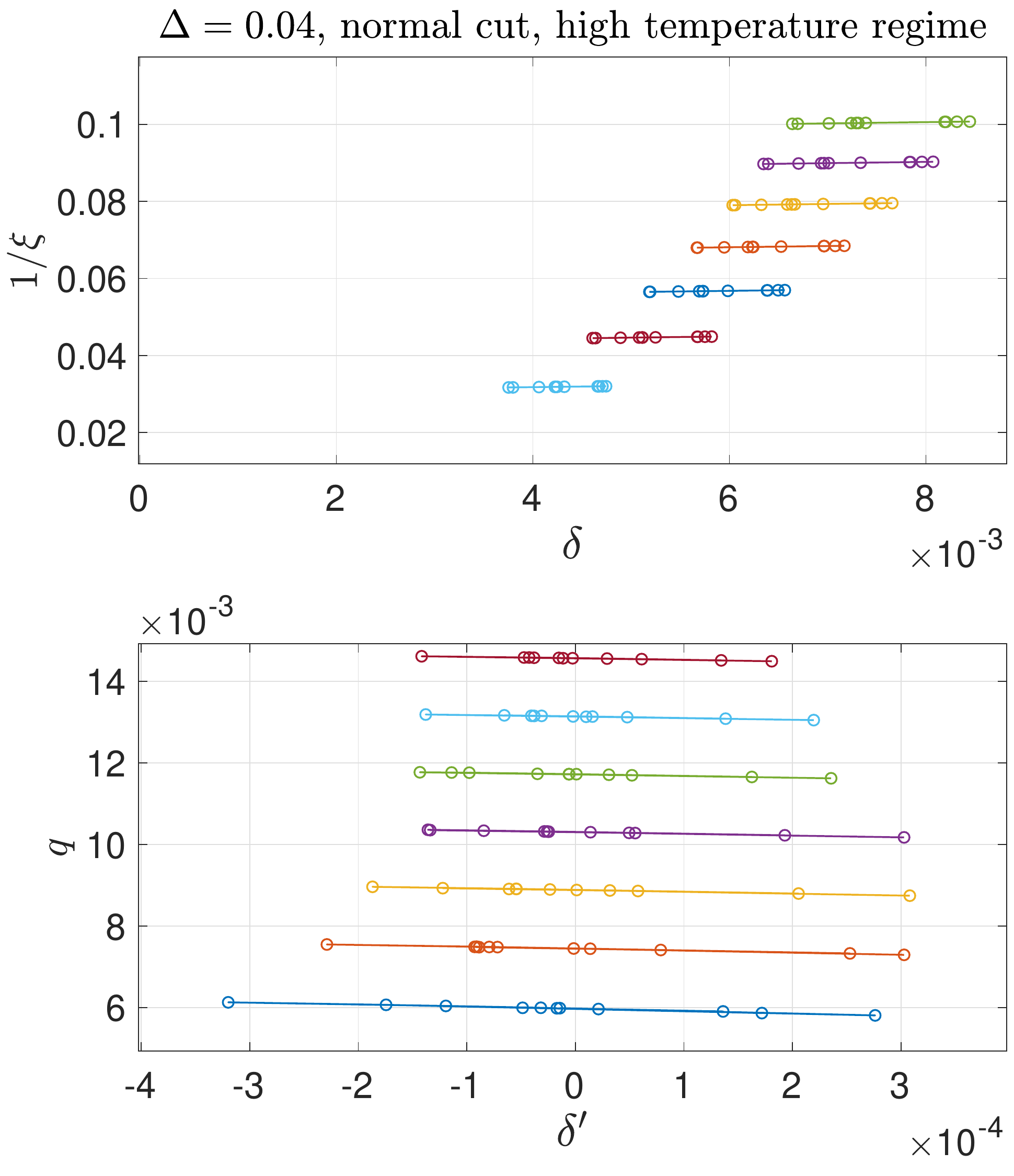}
\caption{Linear extrapolation for both the inverse correlation length (top) and wave vector (bottom) as suggested by Eqs \ref{eq:scaling1}. and \ref{eq:scaling2} with $\chi\in[200,300]$. Each color represents a different temperature.}
\label{fig:extrapolation}
\end{figure}

The accuracy of the results is controlled by the parameter $\chi$. Results would be exact in the $\chi\rightarrow \infty$ limit, and an empirical way to get estimates of physical observables consists in extrapolating results in $1/\chi$. However, it has been suggested\cite{czarnik2018} that a smoother, essentially linear scaling can be obtained for the inverse correlation length with respect to the difference between higher eigenvalues of the transfer matrix, $\delta=\epsilon_i- \epsilon_j, (i,j\neq1)$. We noticed that the wave vector also scales linearly with the difference of phases of higher eigenvalues, $\delta'=\phi_i-\phi_j, (i,j\neq1 )$ (Fig. \ref{fig:extrapolation}). As the bond dimension $\chi$ goes to infinity, one expects all these differences to go to zero. This is the scaling we have used throughout the paper to take the infinite $\chi$ limit for the correlation length and the wave vector.
\begin{align}
  & \frac{1}{\xi(\chi)} = \frac{1}{\xi_{\rm exact}} + b \delta (\chi) \label{eq:scaling1} \\
  & q(\chi)= q_{\rm exact} + b'\delta'(\chi) \label{eq:scaling2}
\end{align}
In practice, we chose 
\begin{equation}
 \delta= \epsilon_4- \epsilon_2, \qquad  \delta' = \phi_4- \phi_2 
\end{equation}
at high temperature, and 
\begin{equation}
\delta= \epsilon_7- \epsilon_4 \qquad 
\end{equation}
at low temperature.

By keeping track of the energy difference $\delta E$ between two iterations, we say that the algorithm converges up to some precision $\delta_E$ if there is a value of $N$ such that $\delta E_N< \delta_E$ and $\delta E_{n+1} < \delta E_{n}$ for every iteration $n<N$. The larger $\chi$, the smaller the reachable precision $\delta_E$. To find out which precision was necessary, we looked at the Potts model. If $\delta_E$ is too large, the effective exponent $\nu(t)$ shows discontinuities. For all simulations used to compute the effective exponents $\nu$ and $\bar{\beta}$ we imposed $\delta_E \simeq o(10^{-9})$. In order to achieve such a precision, all the simulations were done with a minimum bond dimension $\chi=200$. Even with such a bond dimension, some parts of the phase diagram are not accessible to the algorithm. In particular, if the point in the $(T,\Delta)$ plane is too close to the transition, the algorithm is unable to reach the required precision. For example, along the chiral transition line, the energy only converges up to $10^{-4} /10^{-5}$ for typical values of $\Delta$. Furthermore, when the chirality increases, the distance from the transition at which the algorithm converges to $\delta_E \simeq o(10^{-9})$ increases. This might be due to the growing asymmetry of the tensors. At the Potts point, all tensors are symmetric, and the algorithm converges to $\delta_E \simeq o(10^{-9})$ even at the critical temperature. Note also that the convergence of the CTMRG algorithm depends on the choice of isometries. The isometries of Eq. \ref{eq:isometries}  turned out to be a good compromise between computer time and convergence.

Furthermore, we have only considered values of $\xi$ and of $q$ for which the scaling with $\delta$ or $\delta'$ (see Fig. \ref{fig:extrapolation}) was linear to a very good approximation. This has allowed us to reach maximum values ranging between $60$ and $30$ for $\xi_y$ depending on the value of $\Delta$, and significantly smaller values for $\xi_x$ for large $\Delta$, of order $10$ at $\Delta=0.17$. For $q$, very small values could be reached for small $\Delta$ (for instance $q=0.0018$ at $\Delta=0.02$), but the minimum accessible value increased significantly with $\Delta$ to reach $q=0.034$ at $\Delta=0.17$.

\subsection{Effective exponents and Potts point}

\begin{figure}[t!]
\includegraphics[width=0.45\textwidth]{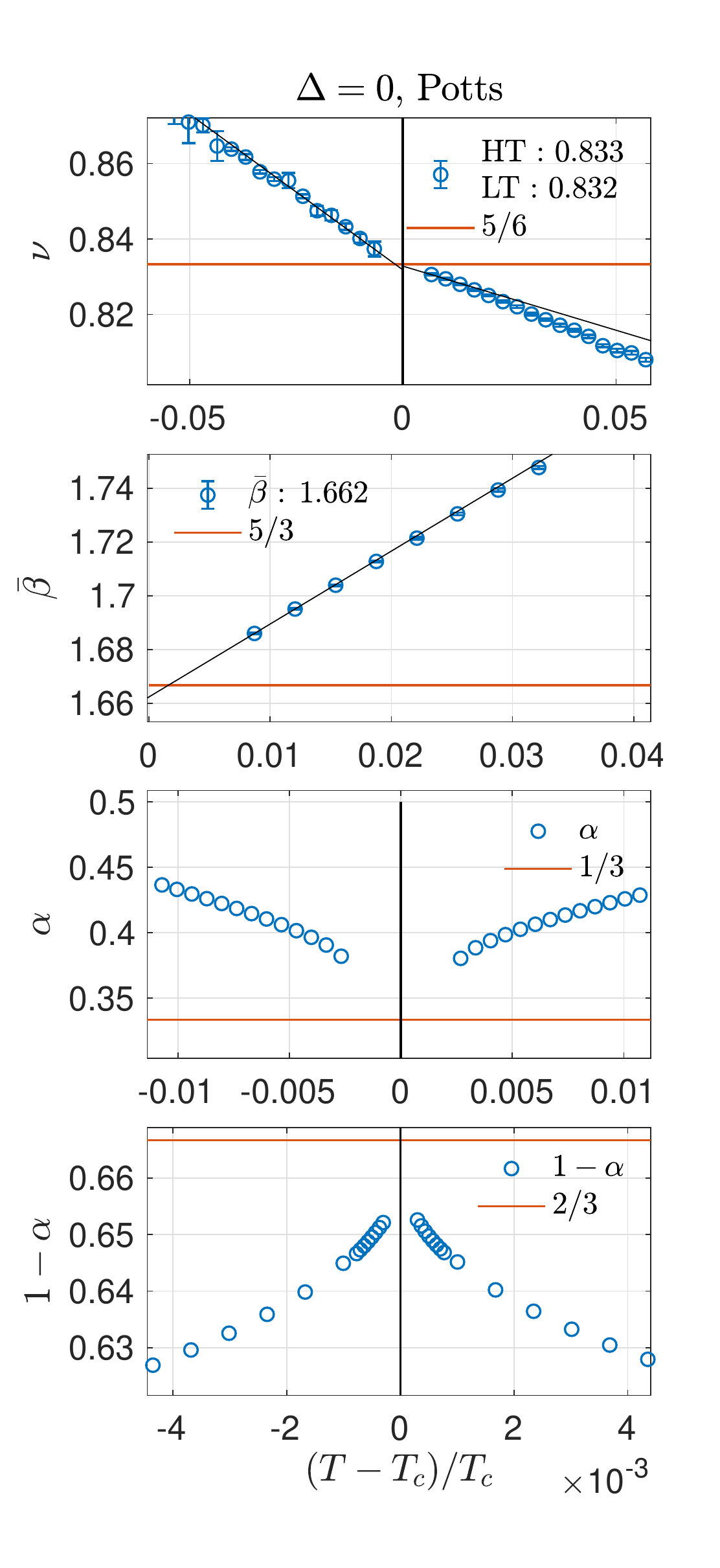}
\caption{Effective exponents of the Potts model: (a) Correlation length $\nu$; (b) Incommensurability $\bar \beta$; (c) Specific heat $\alpha$; (d) Energy per site $1-\alpha$. In all cases the numerical data are consistent with the exact values $\nu=5/6$, $\bar \beta=5/3$, and $\alpha=1/3$ in the limit $T\rightarrow T_c$, but with significant corrections to scaling, especially for $\alpha$. The values quoted in the top two panels are linear extrapolations based on the last two points. There are no error bars on the exponent $\alpha$ because the energy has converged with respect to $\chi$.  We used $\chi=300$.
}
\label{fig:Potts}
\end{figure}

The precise numerical determination of critical exponents is notoriously difficult because of corrections to scaling that severely limit the temperature range where the actual critical exponents can be observed. If the numerical data are precise enough however, and if the critical temperature is known with sufficient precision, a very convenient way of dealing with this problem consists in analyzing the limiting behaviour of effective exponents\cite{pleimling_selke}. Suppose that a quantity $A$ is expected to diverge as $A \propto |t|^{-\theta}$. Then, if one defines an effective exponent $\theta(|t|)$ by
\begin{equation}
\theta(|t|)= -\frac{d \ln A}{d \ln |t|}.
\label{eq:theta_effective}
\end{equation}
the true exponent can be obtained as
\begin{equation}
\theta = \lim_{|t|\rightarrow 0} \theta(|t|)
\label{eq:lim_theta}
\end{equation}
To get a feeling for how important these corrections are, and to benchmark our simulations, let us consider the Potts point $\Delta=0$. At that point, the exponents are known exactly\cite{baxter1980,Baxter_Pearce} and are given by $\nu=5/6$ (correlation length in both directions), $\bar \beta=5/3$ (deviation from commensurability in the disordered phase), and $\alpha=1/3$. Since the critical temperature is known exactly, the only condition to use Eqs. \ref{eq:theta_effective},\ref{eq:lim_theta} to get these critical exponents is to have precise enough data for the correlation length $\xi$, the deviation from commensurability $q$, and the specific heat $C$. This condition is necessary to get small enough error bars when approximating Eq.\ref{eq:theta_effective} by the slope between two consecutive points. The results are shown in Figs.\ref{fig:Potts}(a,b,c,d). The exponent $\bar \beta$ has been obtained by tracking the incommensurability along the line $\Delta=T-T_c$ that terminates at the Potts point $\Delta=0$ and $T=T_c$  and not along the line $\Delta=0$ since the system remains commensurate in the disordered phase along this line. As one can see, the error bars are in most cases very small and in any case always small enough to study the limiting behaviour of the effective exponents. The results for $\nu$ and $\bar \beta$ nicely extrapolate to the true exponents with an accuracy better than $10^{-2}$, but even for these favourable cases corrections to scaling are important, and extrapolating the effective exponent is necessary to get an accurate value. For $\alpha$, the corrections to scaling are very strong, as already known from Monte Carlo simulations\cite{sokal1997}, and even for a reduced temperature as small as 0.0025, the lowest for which we could get precise enough values, the effective exponent is still equal to 0.38, quite far from 1/3. One can do much better however by considering the energy per site $e$, which is expected to have a singularity at the critical temperature of the form $e-e_c\propto |t|^{1-\alpha}$, where $e_c$ is the energy per site at the critical temperature. For the Potts model, its exact value is known. With the notation of Eq.\ref{eq:model}, it is given by $e_c=-(1+\sqrt{3})/2$. There are two main advantages of using the energy instead of the specific heat. First of all, additive corrections to scaling, which are expected to be present, are amplified by taking the derivative, and are thus bigger for the specific heat. 
\footnote{Indeed, assuming that $e-e_c\propto t^{1-\alpha}(1+at^\theta)$ leads to $C\propto t^{-\alpha}(1+a(1+\theta/(1-\alpha))t^\theta)$ so that the coefficient of the correction to scaling is increased by a factor $1+\theta/(1-\alpha)$.}
Besides, for the energy, we can get precise enough data to define the effective exponent much closer to $T_c$ because we do not have to take the numerical derivative. Altogether, the results are much more precise, and for the point closest to $T_c$, the effective exponent we could get for $1-\alpha$ is equal to 0.653, implying a value of 0.347 for $\alpha$, much closer to 1/3. Note also that the effective exponent is still changing, and that its behaviour is consistent with the expected limit $1/3$. Actually, the form of the additive correction to scaling is known exactly\cite{nienhuis1982}, with an exponent 2/3, implying an infinite slope for the effective exponent, consistent with our results. Trying to fit our data leads however to a smaller exponent for the correction to scaling, of the order of 0.45, an effect already observed in Monte Carlo simulations\cite{sokal1997} and attributed to higher-order corrections to scaling.

\subsection{Critical temperature}

The analysis in terms of effective exponents is only possible if the critical temperature is known with sufficient accuracy. To be more precise, since we can often reach reduced temperatures $|t|$ as small as $0.01$ or even smaller, the analysis of the effective exponents requires to know the critical temperature with a precision of about $10^{-4}$. In that respect, the standard way to proceed, which consists in using both the critical temperature and the exponent as adjustable parameters in a fit performed over a finite temperature window, is completely inadequate because fitting different quantities in this way leads to estimates of the critical temperature that can vary by as much as $10^{-2}$. So, to determine $T_c$, we chose to constraint the fit by physical considerations. The strategy we have adopted is the following: 

(i) Far enough from Potts, i.e. for large enough $\Delta$, it is clear from previous approaches that the transition has to be a two-step process, with, coming from low temperature, a Pokrovsky-Talapov transition followed by a Kosterlitz-Thouless transition. The critical properties of the PT transition are well known (see below). In particular, coming from low temperature, the correlation length in the $y$ direction is expected to diverge with an exponent $\nu_y=1$. At the same time, coming from high temperature, the correlation length is expected to diverge exponentially at a higher temperature, leading to an effective exponent that would seem to diverge if calculated using the PT transition temperature. For large $\Delta$, we have thus determined the Pokrovsky-Talapov critical temperature $T_{PT}$ but adjusting it so that the effective exponent $\nu_y$ coming from the commensurate phase tends to 1 at the critical point. This criterion fixes the critical temperature to within $10^{-4}$.

(ii) For small $\Delta$, this analysis breaks down because the correlation length in the high temperature phase does not diverge any more before that of the low temperature phase. In that case, the only alternative is that there is a single continuous transition, as advocated by Huse and Fisher. So, to determine the critical temperature, we have looked for the temperature such that the effective exponents $\nu_y$ from high and low temperature are consistent with each other. This turns out to be an extremely stringent condition because temperatures outside a very narrow window lead to opposite behaviours of the effective exponents, with one becoming very large and the other one becoming very small. As we shall see, this again fixes the critical temperature within $10^{-4}$.

\subsection{Error bars}
The effective exponent for the correlation length can be written as
\begin{align}
 \nu=\frac{t}{\xi}\frac{\partial \xi}{\partial t}.
\end{align}
 With a central finite difference, the discretization of the first derivative gives
\begin{align*}
    \nu= \frac{t}{\xi}\frac{\xi(t+h)-\xi(t-h)}{2h} + o(h^2).
\end{align*}
Then, from a Taylor expansion to first order, one obtains
\begin{align}
     \nu(t+\delta t, \xi + \delta \xi)& = \nu +\delta \nu
\end{align}
with
\begin{align*}
    \delta \nu = \nu \left( \frac{\delta T_c}{|T-T_c|} + \frac{\delta \xi(t)}{\xi} + \frac{\delta \xi(t+h) + \delta \xi (t-h) }{\xi(t+h)-\xi(T+h)}  \right) 
\end{align*}
The same calculation can be done for $\bar{\beta}$.
The effective exponents have two different sources of errors. The main one, which becomes more important close to criticality, comes from the uncertainty of the critical temperature, while the second one comes from the accuracy of the linear fit in the extrapolation of the inverse correlation length and wave vector. 

For $\alpha$, $\nu_x$ and $\bar{\beta}$, the error bars from $\delta T_c$ represent what those exponents would have been if one had considered $T_c \pm \delta T_c$ as critical temperature. As discussed above, the critical temperature for both the chiral and Pokrovsky-Talapov transitions has been determined with $\nu_y$. Therefore we do not take into account the source of error from $\delta T_c$ for the error bars of $\nu_y$.

\section{Results}

Let us now turn to the main results of this paper. We will scan the phase diagram starting from large $\Delta$, where a PT transition can be fully characterized, to smaller values of $\Delta$, where there appears to be a unique transition with universal exponents, through a Lifshitz point where the intermediate critical phase disappears.

\subsection{Pokrovsky-Talapov transition}

The Pokrovsky-Talapov transition is ubiquitous in statistical and condensed matter physics. It describes a transition from an ordered phase with a finite correlation length to a critical phase with algebraic correlations. For the model of Eq. \ref{eq:model}, the ordered phase is uniform up to $\Delta=1/2$, a range to which we will limit our investigation since the phase diagram can be shown to be symmetric with respect to $\Delta=0 $ with period 1. In that parameter range, the critical phase consists of fluctuating domain walls with an average separation $2 \pi/q$, where the wave vector $q$, which vanishes in the ordered phase, measures the incommensurability. In that phase, the correlations are algebraic. 

This transition is very anisotropic. Coming from the incommensurate phase, it is characterized by $\bar \beta=1/2$, where $\bar \beta$ describes the critical behaviour of $q$ close to the transition: $q\propto t^{\bar \beta}$, and by a specific heat exponent $\alpha=1/2$. Coming from the ordered phase, the correlation lengths $\xi_x$ and $\xi_y$ diverge with exponents $\nu_x=1/2$ and $\nu_y=1$ along $x$ and $y$ respectively, so that the anisotropy exponent, most often referred to as the dynamical exponent by analogy with the physics of quantum models in dimension 1+1, is given by $z=2$. In addition, the specific heat does not diverge. It just has a $|t|/\ln |t|$ singularity.

\begin{figure}[t!]
\includegraphics[width=0.45\textwidth]{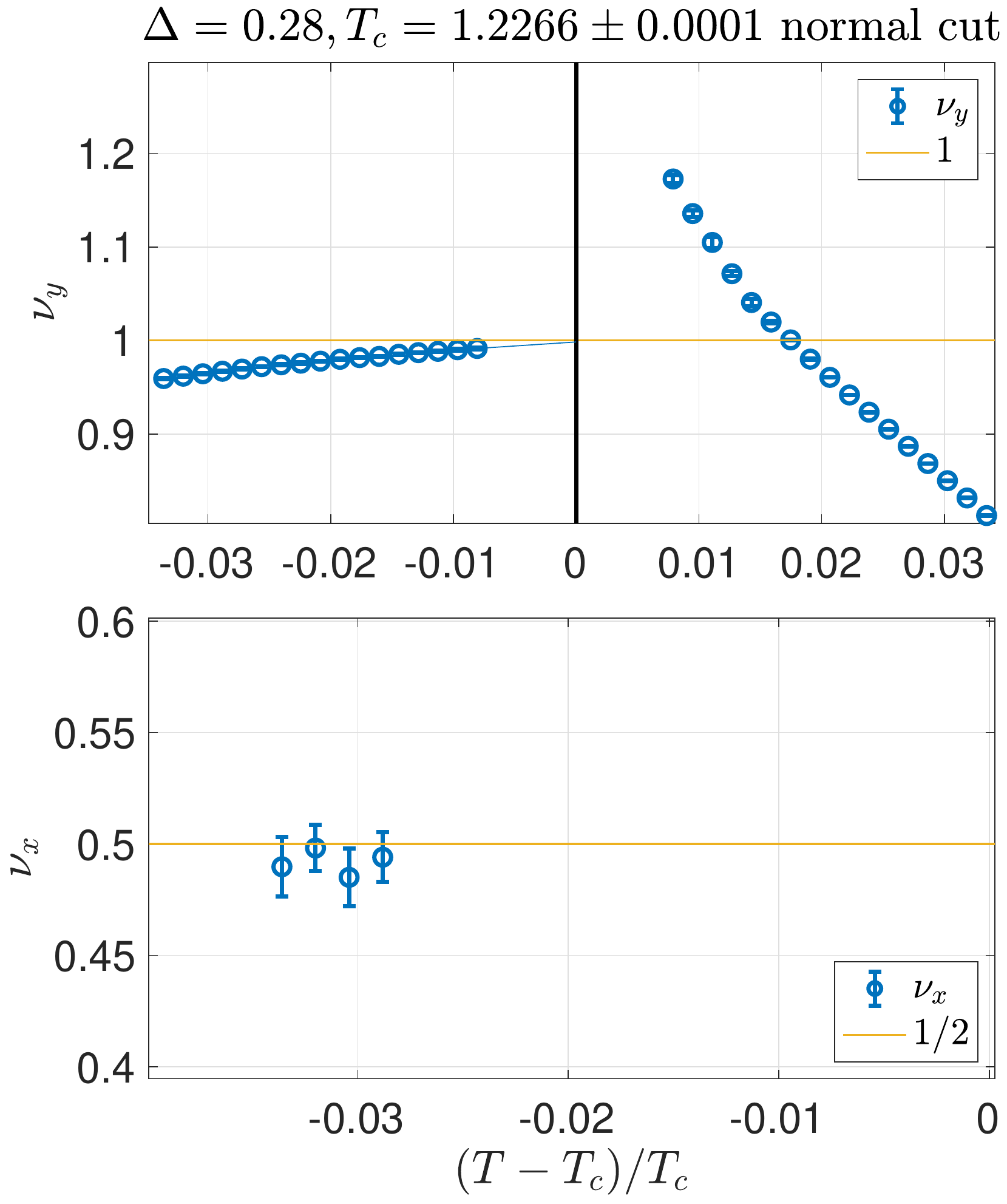}
\caption{Effective exponents $\nu_x$ and $\nu_y$ of the chiral Potts model at the Pokrovsky-Talapov transition for $\Delta=0.28$. The numerical data are consistent with the exact values $\nu_x=1/2$ and $\nu_y=1$. The linear extrapolation of $\nu_y$ is based on the last two points.
}
\label{fig:Delta=0.28}
\end{figure}

The results that we have obtained for the correlation length exponents at $\Delta=0.28$ are summarized in Fig. \ref{fig:Delta=0.28}. Assuming that $\nu_y$ goes to 1 at low temperature to fix the critical temperature $T_{\rm PT}$ leads to an effective exponent for $\nu_y$ at high temperature that is larger than 1 and increases fast upon approaching $T_{\rm PT}$. This implies that this is not a simple order-disorder transition, in which case the exponent should be the same on both sides of the transition, but this is consistent with an infinite correlation length on the high-temperature side of the transition. Let us emphasize that the effective exponent $\nu_y$ estimated at high temperature using $T_{\rm PT}$ is actually meaningless because the correlation length is expected to diverge at a Kosterlitz-Thouless transition with critical temperature $T_{\rm KT}>T_{\rm PT}$. It will nevertheless prove useful to keep track of the divergence of the correlation length at high temperature using this effective exponent because it is expected to tend to the same value as its low temperature counterpart when the intermediate phase disappears.
The correlation length exponent in the other direction $\nu_x$ is also consistent with a PT transition, with a value very close to 1/2. 

\begin{figure}[t!]
\includegraphics[width=0.45\textwidth]{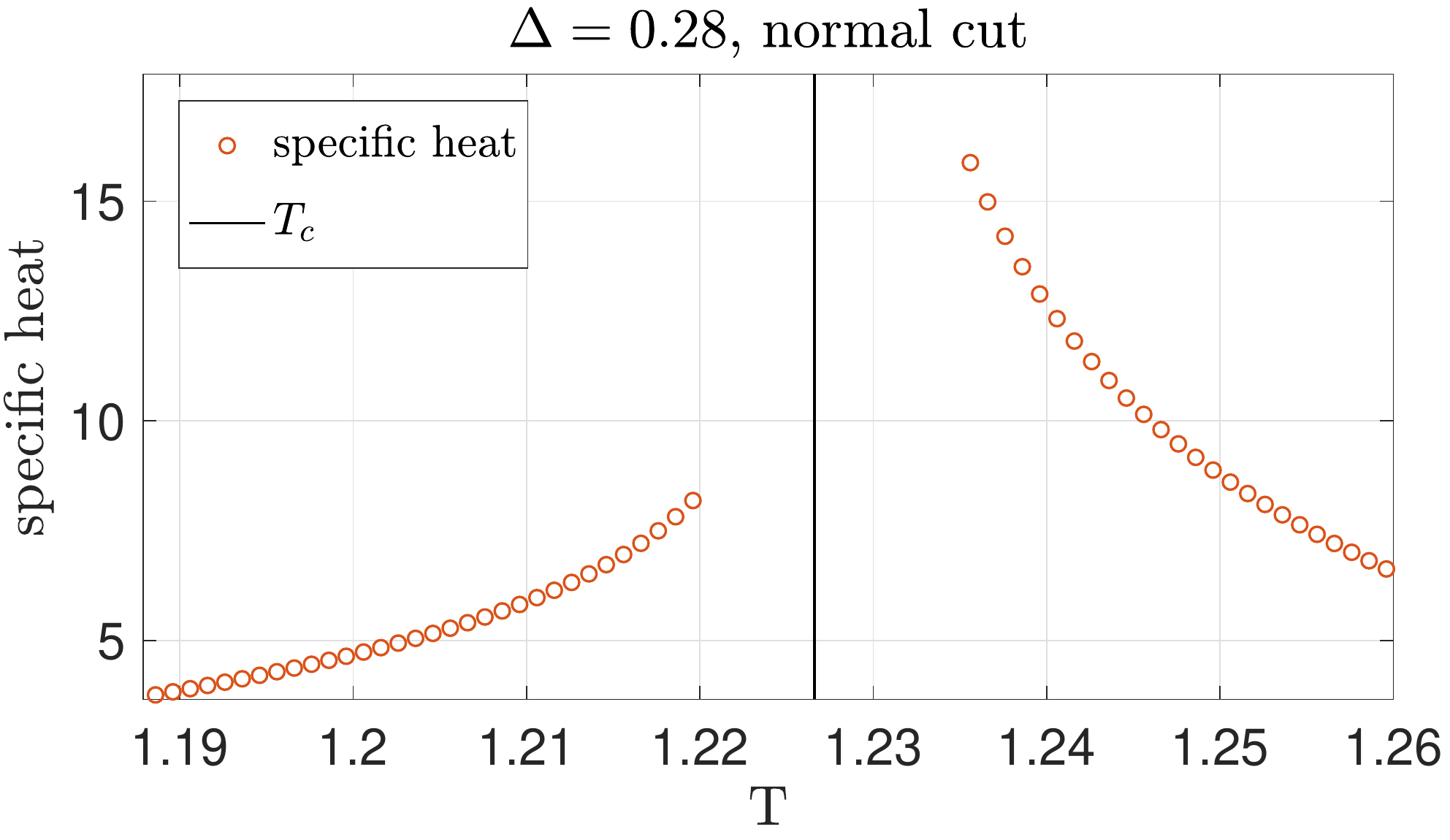}
\caption{Specific heat of the chiral Potts model at $\Delta=0.28$. As expected for a Pokrovsky-Talapov transition, it is very asymmetric.}
\label{fig:specific_heat_PT}
\end{figure}

As a further check of the PT universality class, we have compared the behaviour of the specific heat on both sides of the transition in Fig. \ref{fig:specific_heat_PT}. As expected, it is very asymmetric, with no clear sign of a divergence on the low temperature side.

The critical behaviour inside the critical phase above the PT transition is unfortunately not accessible. For that value of $\Delta$, the critical phase is already extremely narrow. We actually do not have an estimate of $T_{\rm KT}$, but we know that the correlation length is still finite (and not yet very large, $\xi_y\simeq 25$) at $|t|=0.0063$, implying that $T_{\rm KT}-T_{\rm PT}<0.0077$. To access the critical behaviour we should reach temperatures within $10^{-3}$ of $T_{\rm PT}$ or smaller, but our algorithm does not converge so close to $T_{\rm PT}$ in the critical phase.

\subsection{Lifshitz point}

\begin{figure}[t!]
\includegraphics[width=0.45\textwidth]{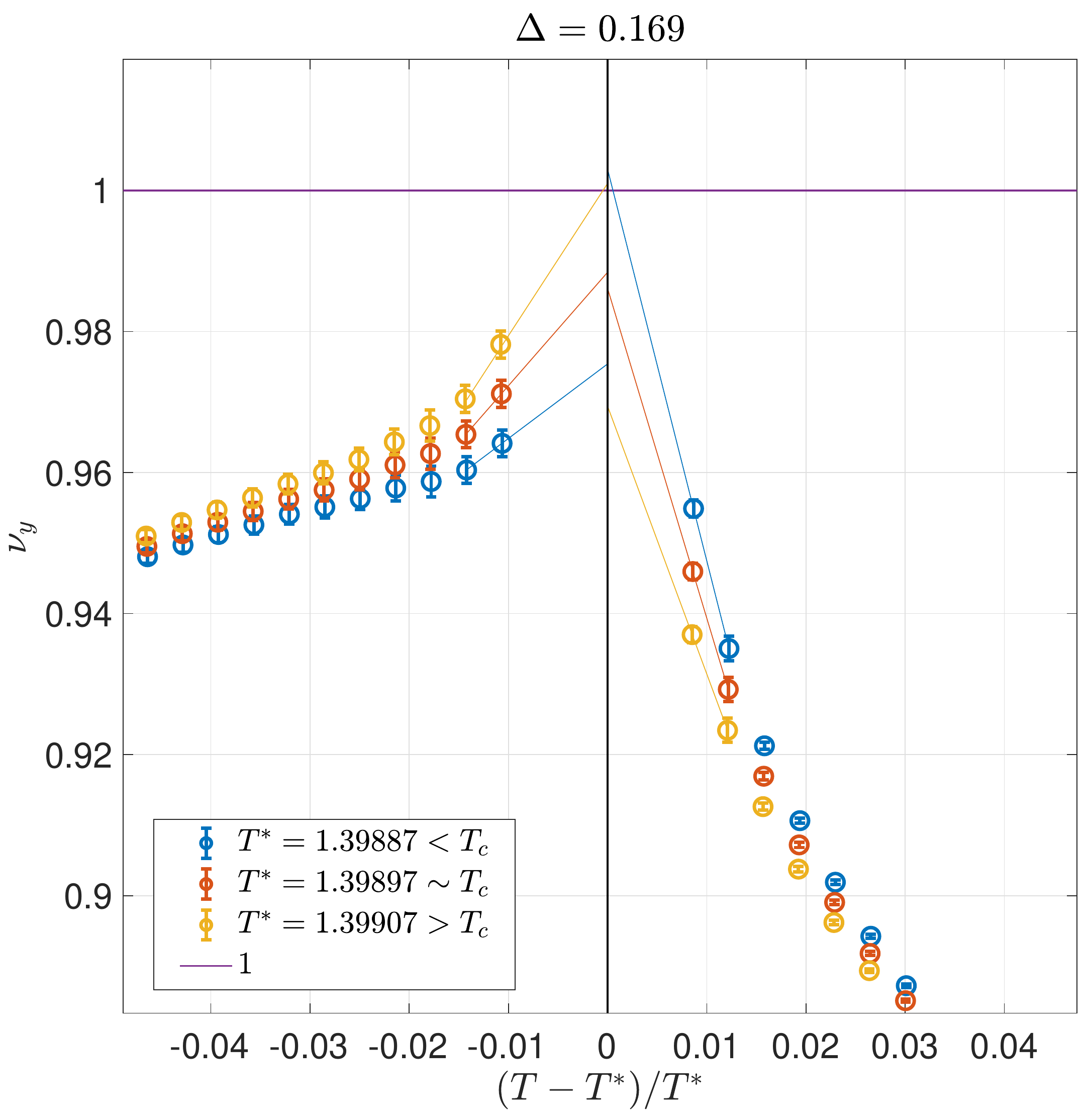}
\caption{Effective exponents $\nu_y$ of the chiral Potts model at $\Delta=0.169$ for three termperatures around the critical temperature. This is the first point coming from high temperature at which the limit of the high temperature exponent is consistent with that of the low temperature one, and we interpret it as a Lifshitz point. The extrapolations are just linear ones based on the last two points. Given the upward curvature of both curves due to corrections to scaling, the high and low temperature effective exponents are consistent with a limiting value $\nu_y=1$. 
}
\label{fig:Delta=0.169}
\end{figure}

Upon reducing $\Delta$, the results evolve smoothly until $\Delta \simeq 0.17$, where the limit of the high temperature effective exponent $\nu_y$ becomes more or less compatible with that of the low-temperature one. Scanning different values of $\Delta$ between 0.165 and 0.17, we located the point where they actually become compatible at $\Delta=0.169$, as shown in Fig. \ref{fig:Delta=0.169}. 

Since the low and high temperature correlation lengths along $y$ diverge at the same temperature, the intermediate critical phase has to disappear at this point, which can thus be identified as a Lifshitz point. Note also that both the high and low temperature effective exponents are consistent with a limiting value $\nu_y=1$. They are severe corrections to scaling however, hence some uncertainty on the location of the Lifshitz point, which we estimate at $\Delta_L=0.169\pm0.003$.

\subsection{Chiral transition}

Below that point, the critical behaviour changes quite drastically. The high and low temperature correlation lengths appear to diverge roughly at the same temperature. As explained in the previous section, a very precise way to locate the critical temperature is to impose that the effective exponents at high and low temperature calculated with the same critical temperature tend to the same value. The resulting exponent is roughly consistent with $\nu_y=1$ just below the Lifshitz point, but it tends to become smaller upon approaching the Potts point. As we shall see below, general scaling relations between exponents suggest that this is a crossover effect, and that this exponent $\nu_y$ remains equal to 1 down to the Potts point, where it jumps to 5/6.
For the time being, we just note that simply imposing consistent exponents fixes the critical temperature to an accuracy of $10^{-4}$. With this accuracy, the discussion of the other exponents in terms of effective exponents becomes meaningful. Let us discuss separately the results we have obtained for the specific heat exponent $\alpha$, the incommensurability exponent $\bar \beta$, and the correlation length exponents $\nu_x$ and $\nu_y$, concentrating on two points, one close to the Potts point, $\Delta=0.04$, the other one close to the Lifshitz point, $\Delta=0.16$.

\subsubsection{Specific heat exponent $\alpha$}

\begin{figure}[t!]
\includegraphics[width=0.49\textwidth]{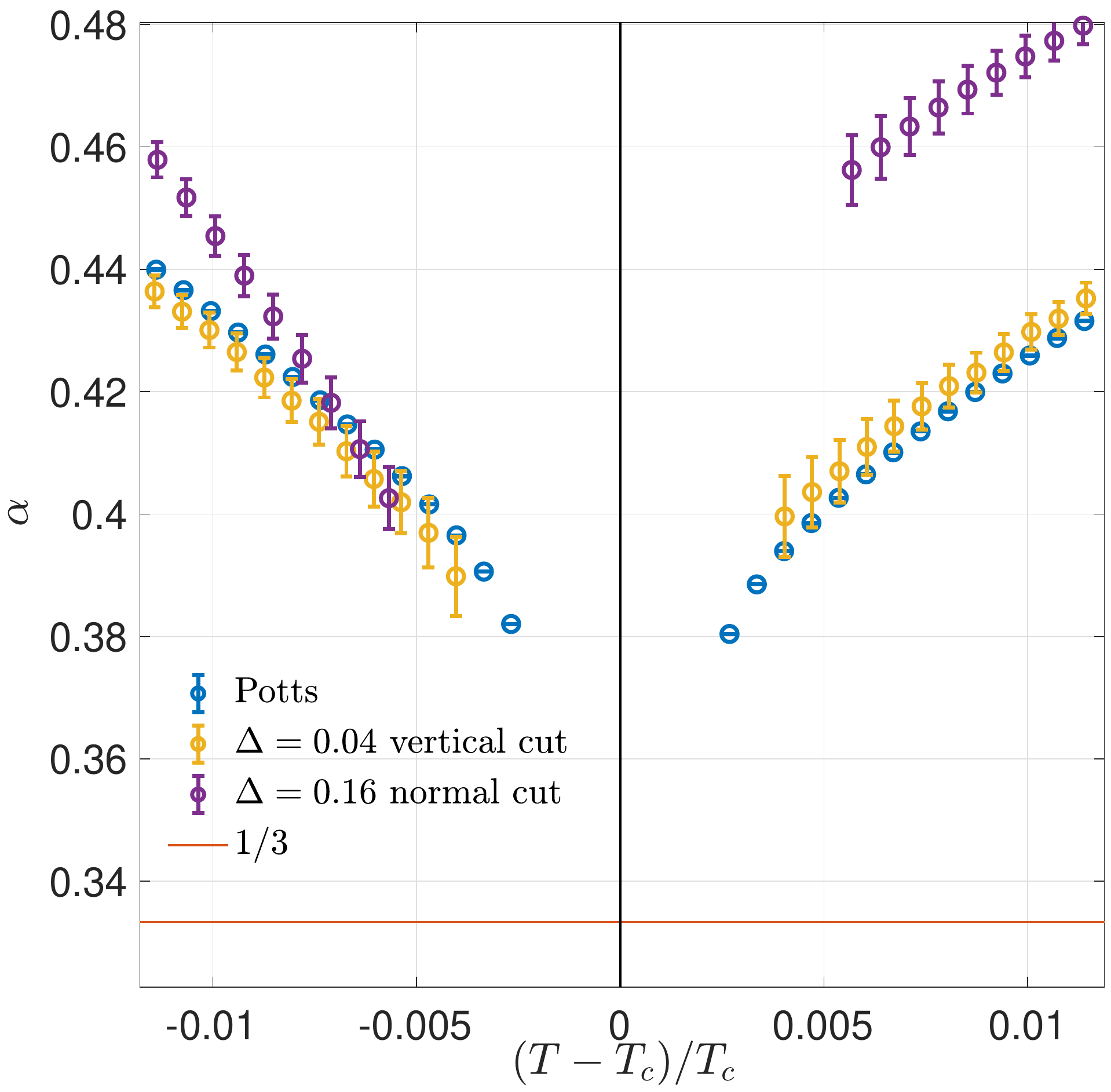}
\caption{Effective exponent $\alpha$ at the Potts point and along the chiral transition (at $\Delta=0.04$ and $\Delta=0.16$) extracted from the specific heat.
}
\label{fig:specific_heat}
\end{figure}

The results for the effective exponents of the specific heat are summarized in Fig.\ref{fig:specific_heat}. For $\Delta=0.04$, the results are essentially identical to those of the Potts model within the error bars, a first indication that the exponent $\alpha$ has not changed much (if at all). However, the situation is much less clear for $\Delta=0.16$, with a very asymmetric behaviour between low and high temperatures. The low temperature data are again consistent with an exponent close to 1/3, but the high temperature results would rather point to an exponent around 0.43.

\begin{figure}[t!]
\includegraphics[width=0.45\textwidth]{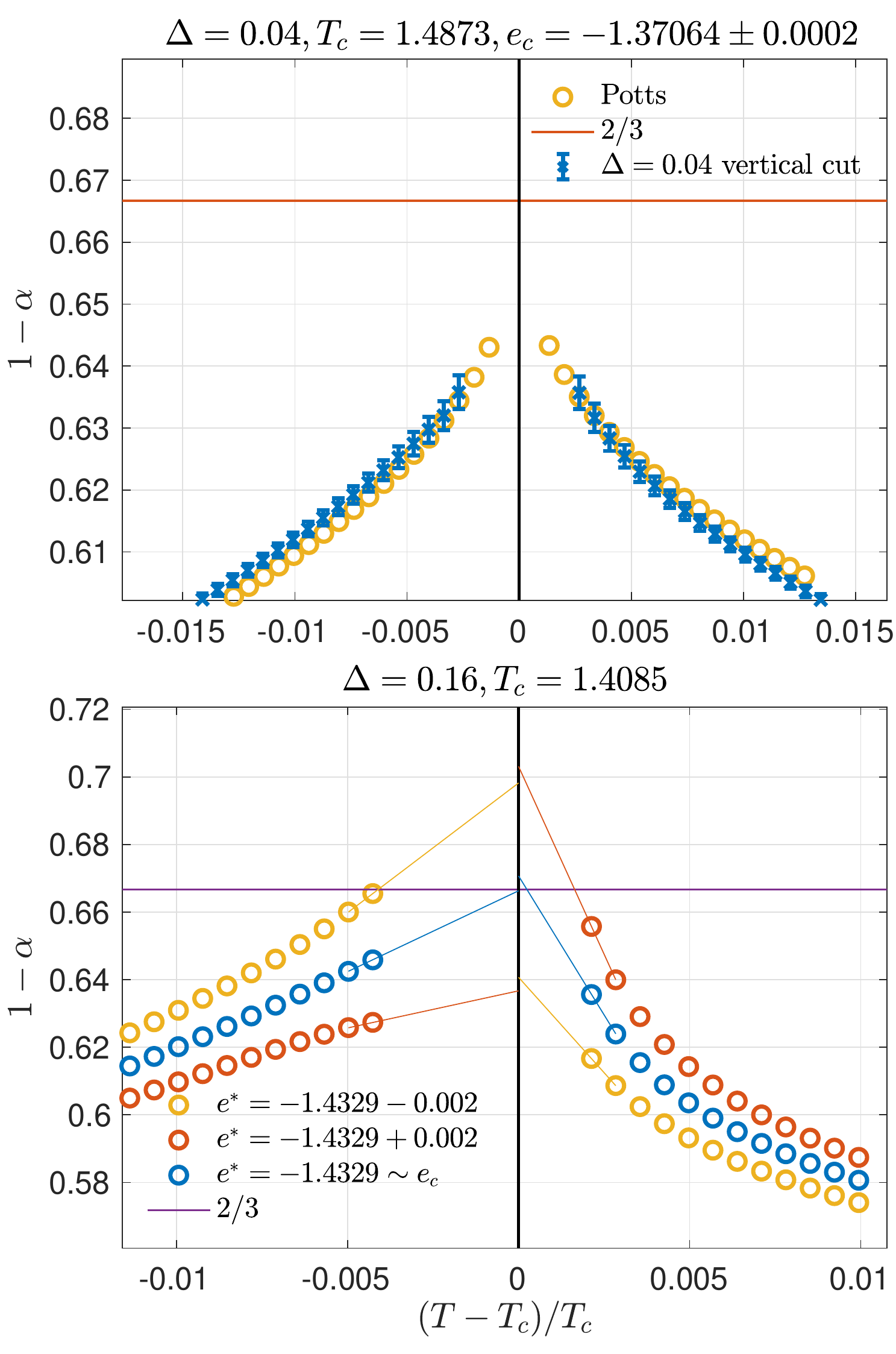}
\caption{Energy effective exponent $1-\alpha$ extracted from the energy along the chiral transition at $\Delta=0.04$ (upper panel) and $\Delta=0.16$ (lower panel). At $\Delta=0.04$, the effective exponent is barely distinguishable from the Potts point values close to the transition. At $\Delta=0.16$, imposing a unique transition gives an extrapolated value close to the Potts exponent 2/3. The extrapolations are linear and based on the last two points. 
}
\label{fig:energy}
\end{figure}

As for the Potts model, one can get much more reliable information by looking at the energy. In keeping with our strategy, we just assume that there is a direct order-disorder continuous phase transition, hence that the critical exponents that describe the non-analyticity of the energy density are the same on both sides of the transition. To ensure this, we use the energy at the critical point $e_c$ as an adjustable parameter. Let us emphasize that we do not make any assumption on the value of the exponent. The results are shown in Fig.\ref{fig:energy}. Quite remarkably, if we impose that the high and low-temperature effective exponents converge to the same value, the resulting estimate is consistent with 2/3, hence with a value of $\alpha=1/3$, for both $\Delta=0.04$ and $\Delta=0.16$.

\subsubsection{Incommensurability exponent $\bar \beta$}

To get a meaningful estimate of the exponent $\bar \beta$, we must have access to values of $q$ small enough to see at least the beginning of the critical behaviour. This turns out to be possible close to the Potts point, but by the time $\Delta$ reaches 0.16, the constant $q$ lines are very dense, and we are limited to $q$ values larger than 0.05 for which the behaviour is essentially linear and does not reveal the critical behaviour. 

\begin{figure}[t!]
\includegraphics[width=0.45\textwidth]{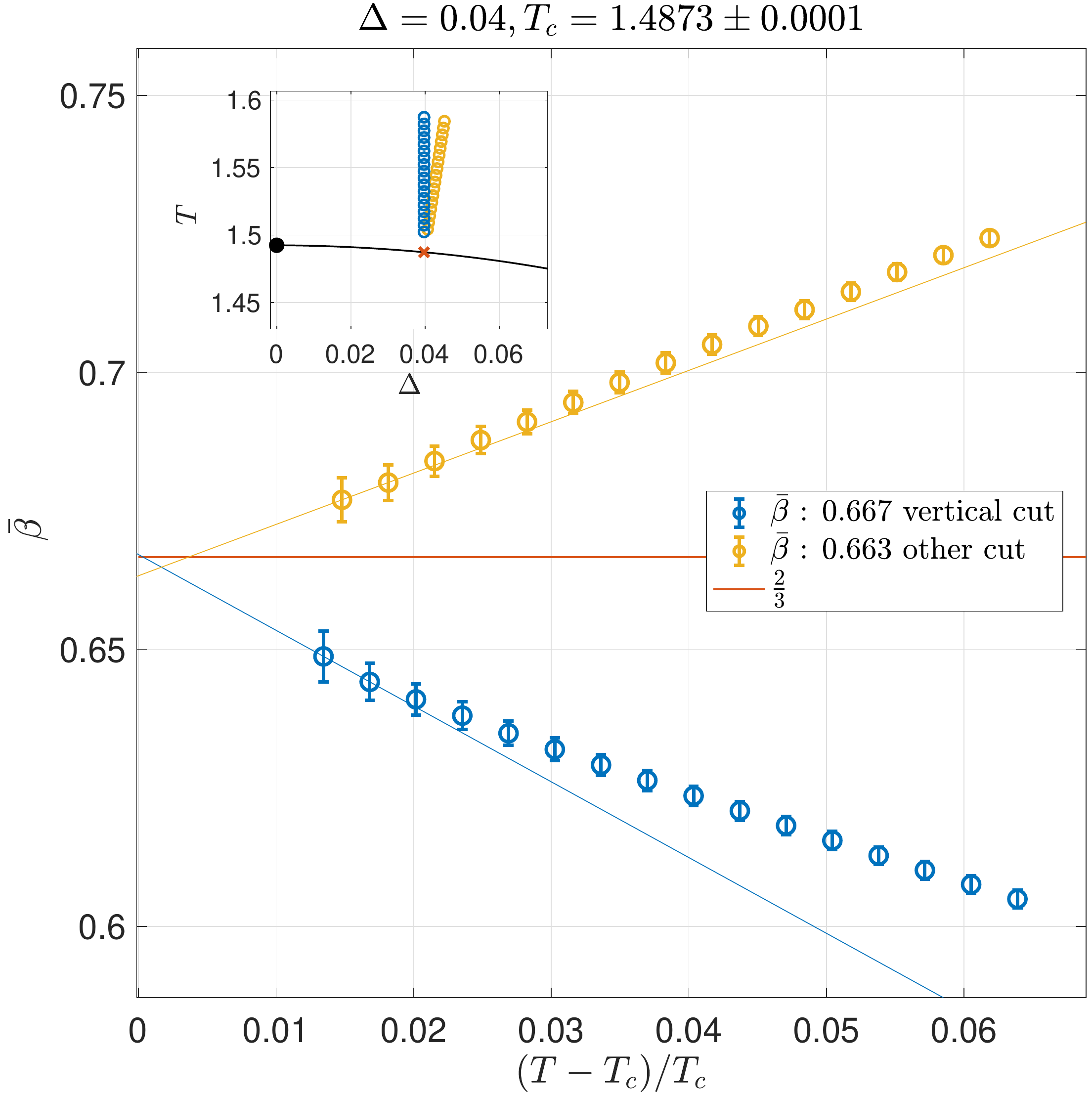}
\caption{Effective $\bar \beta$ exponent of chiral Potts model at $\Delta=0.04$, close to the Potts point, along two cuts shown in the inset. The values quoted for $\bar \beta$ are linear extrapolations based on the last two points.
}
\label{fig:beta0.04}
\end{figure}

However, for small $\Delta$, we can reach much smaller values of $q$, and the critical behaviour becomes accessible. The results for the exponent $\bar \beta$ at $\Delta=0.04$ along two cuts are shown in Fig. \ref{fig:beta0.04}. The smallest $q$ that we could reach was 0.005. These results point to a value $\bar \beta$ between 0.663 and 0.667, much smaller than the value $\bar \beta=5/3$ expected (and confirmed numerically) at the Potts point. Similar results have been obtained for larger $\Delta$ as long as the critical behaviour could be accessed.

\subsubsection{Correlation length exponents $\nu_x$ and $\nu_y$}

\begin{figure}[t!]
\includegraphics[width=0.45\textwidth]{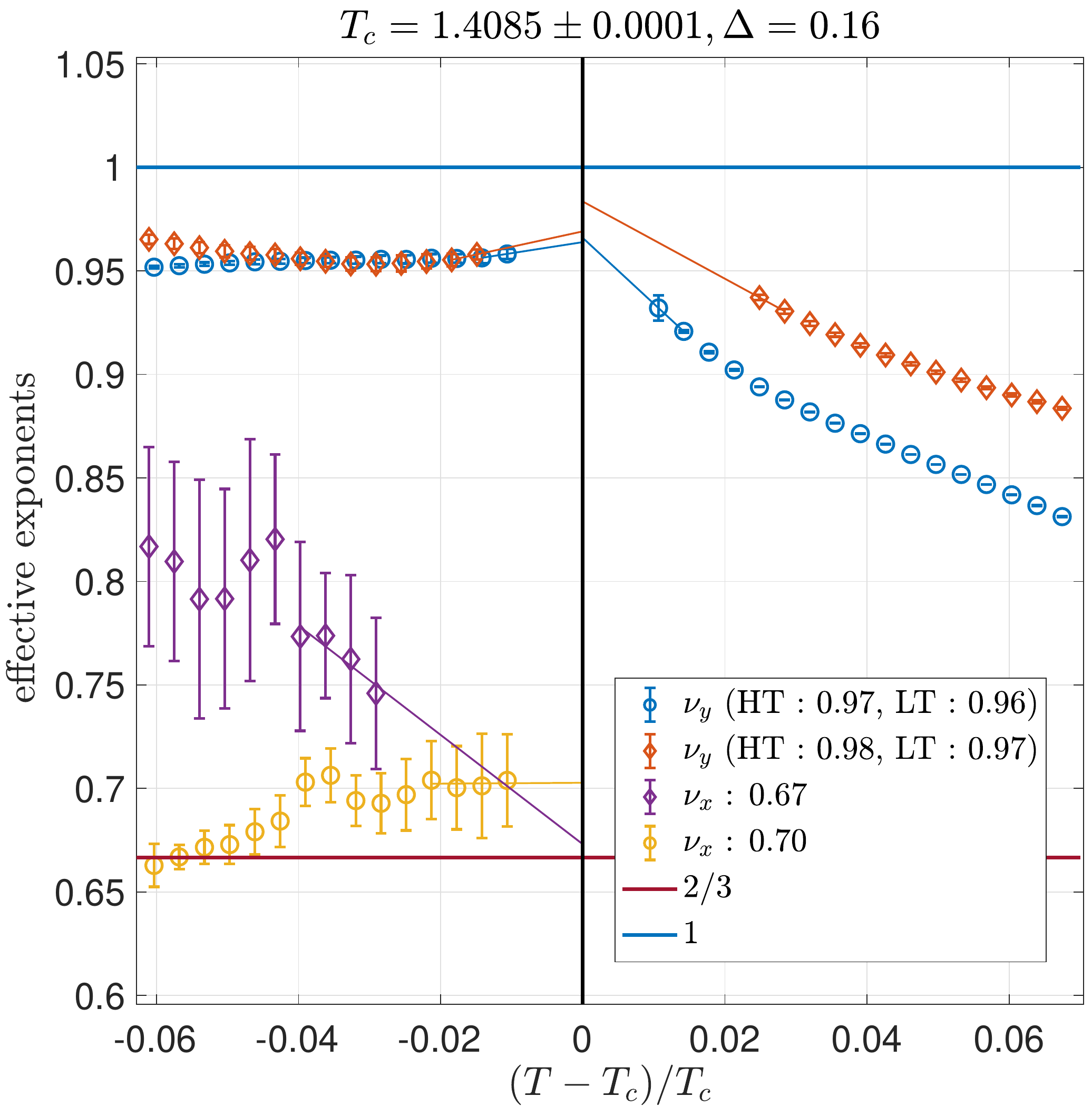}
\caption{Correlation length effective exponents of the chiral Potts model at  $\Delta=0.16$, where the transition is a direct one and is expected to be in the Huse-Fisher universality class. Note that the exponents have been obtained along two different cuts, a vertical one (red and purple diamonds), and a cut normal to the critical temperature line in the $T-\Delta$ plane (yellow and blue circles). The corrections to scaling are in most cases rather different along the two cuts, but the limiting values point to exponents $\nu_y\simeq 0.97$ and $\nu_x\simeq 0.685$. The lines 1 and 2/3 correspond to the exponents of the self-dual chiral Potts model. The values quoted for the exponents are linear extrapolations based on the last two points for $\nu_y$ and on the last four points for $\nu_x$.
}
\label{fig:Delta=0.16}
\end{figure}

The results for the correlation length exponents are very different for small and large $\Delta$. For small $\Delta$, they depart only mildly from the Potts value 5/6 (see examples below for $\nu_y$ in Fig.\ref{fig:crossover}). However, upon approaching the Lifshitz point, the scaling becomes very anisotropic, as demonstrated for $\Delta=0.16$ by the
results shown in Fig. \ref{fig:Delta=0.16}. The exponents have been obtained along two different cuts, a vertical one, and a cut normal to the critical temperature line in the $T-\Delta$ plane. The corrections to scaling are clearly very large, especially for $\nu_x$, but a consistent picture nevertheless emerges from simple linear extrapolations, with $\nu_y=0.97\pm0.04$ and $\nu_x=0.685\pm0.06$. Note that given the upward curvature of $\nu_y$, the value of $\nu_y$ is probably slightly underestimated. Rather than trying more refined fits, for which we lack a theoretical basis, we have adopted conservative error bars corresponding to twice the difference between the smallest and highest estimates. The scaling is in any case very anisotropic, with a dynamical exponent $z\simeq1.42\pm0.2$.

\section{Discussion}

The picture that emerges from our results for the commensurate-incommensurate transition of the chiral three-state Potts model is that of a continuous order-disorder transition from the Potts point at $\Delta=0$ to a Lifshitz point at $\Delta=0.169(3)$, followed by a Pokrovsky-Talapov transition into a critical phase. It is summarized in Fig. \ref{fig:phase_diagram}. The extent of the critical phase above the PT transition is purely indicative, but the evidence of such a phase, hence of a KT transition, is clear because the correlation length diverges before the PT transition when coming from the high-temperature phase. 

\subsection{Huse-Fisher chiral universality class}

Between the Potts point and the Lifshitz point, the transition is a priori characterized by 4 exponents: $\nu_x$, $\nu_y$, $\bar \beta$, and $\alpha$. These exponents are not independent however. First of all, they are expected to be related by hyperscaling:
\begin{equation}
\nu_x+\nu_y=2-\alpha
\label{eq:hyper_scaling}
\end{equation}
Besides, as emphasized by Huse and Fisher, the transition cannot be in the Potts universality class for $\Delta>0$ because the chirality introduced by $\Delta$ is relevant at the Potts point, and if it is not a Pokrovsky-Talapov transition, the exponent $\nu_x$ and $\bar \beta$ are expected to be equal:
\begin{equation}
\bar \beta=\nu_x
\label{eq:huse_fisher}
\end{equation}

Although the scaling is very anisotropic, the first relation is approximately satisfied for $\Delta=0.16$. Indeed, the exponents $\nu_y\simeq 0.97$ and $\nu_x\simeq 0.685$ lead to $\nu_x+\nu_y\simeq 1.655$, in good agreement with $2-\alpha$ if $\alpha\simeq 1/3$, as suggested by our numerical data. 
Unfortunately $\bar \beta$ could not be measured at that point.

Close to the Potts point, the situation is more subtle. Since the apparent values of $\nu_x$, $\nu_y$, and $\alpha$ are the same as at the Potts point, hyperscaling is of course satisfied, but the relation $\bar \beta=\nu_x$ is badly violated since $\bar \beta \simeq 2/3$ while $\nu_x \simeq 5/6$. So the estimation of at least one of the exponents $\bar \beta$ or $\nu_x$ has to be wrong, presumably because of very severe crossover effects (see below). So, since one cannot get arbitrarily close to the critical temperature, one can expect to see the Potts exponents for small $\Delta$. This is precisely what happens for $\nu_x$ and $\nu_y$, but not for $\bar \beta$. At the Potts point, we got a value of $\bar \beta$ perfectly consistent with the exact value $5/3$, but for $\Delta$ as small as 0.04, we already got a completely different value of $\bar \beta$ very close to 2/3. The conclusion that imposes itself is that $\bar \beta$ is apparently not affected by the proximity of the Potts point, while $\nu_x$ and $\nu_y$ are, and that it is the value of $\bar \beta$ that should be trusted. This conclusion is actually supported by the evolution of the results upon increasing $\Delta$: $\nu_x$ and $\nu_y$ change continuously upon increasing $\Delta$, and $\nu_y$ approaches rapidly a value close to 1, while $\nu_x$ decreases to become consistent with $\bar \beta$, which does not change significantly as long as it can be meaningfully measured.

At the same time, the exponent $\alpha$ of the specific heat does not seem to depart at all from the Potts value 1/3 up to the Lifshitz point.
So if we assume that $\alpha=1/3$ for $\Delta=0.04$, and that $\bar \beta \simeq 2/3$, we get $\nu_x\simeq 2/3$ from Eq. \ref{eq:huse_fisher}, and then $\nu_y\simeq 1$ from hyperscaling. Within the error bars, these values are fully consistent with those obtained at $\Delta=0.16$, close to the Lifshitz point. So the numerical evidence points to a universality class with exponents that do not change along the chiral transition. Assuming this to be the case, the most precise estimates are those of $\bar \beta =0.665\pm0.008$ at $\Delta=0.04$, and of $\nu_y=0.98\pm0.04$ at $\Delta=0.169$.

Quite remarkably, there is a generalization of the Potts model with asymmetries in both directions $\Delta_x$ and $\Delta_y$ for which these exponents are known exactly, namely the self-dual case $\Delta_x = \pm i \Delta_y$, which is integrable for $\Delta_x=\pi/6$, and which has been solved by Cardy using renormalization group arguments for the chiral term treated as a perturbation of the conformal field theory that describes the three state Potts model\cite{cardy}. For that model, $\alpha=1/3$, and $\nu$ takes two different values respectively equal to 2/3 and 1 along two directions that rotate upon increasing $\Delta_x$. This model is of course very different from the model of Eq. \ref{eq:model}. In particular it has complex Boltzmann weights. Still it is quite remarkable that the critical exponents that emerge from the present numerical study are consistent with these exact results, and it is tempting to speculate that the critical exponents of the chiral transition of the model of Eq. \ref{eq:model} are exactly given by $\nu_x=\bar \beta = 2/3$, $\nu_y=1$, and $\alpha=1/3$. 

Finally, let us comment on the dynamical exponent $z$, also called the anisotropy exponent, and defined as $z=\nu_y/\nu_x$. Our identification of a unique universality class with exponents that do not evolve as a function of $\Delta$ point to a dynamical exponent that does not vary either, and that is pinned at the value $z=3/2$ along the chiral boundary, a value intermediate between $z=1$ at the Potts point and $z=2$ along the Pokrovsky-Talapov transition.

\subsection{Crossovers and previous investigations}

The numerical investigation of the model of Eq. \ref{eq:model} has a long history. Several numerical approaches have been used ranging from Monte Carlo\cite{Selke1982,houlrik1986} to finite-size renormalization group\cite{Duxbury,yeomans1985}, finite-size transfer matrix\cite{bartelt,stella1987,Everts_1989} and more recently to DMRG\cite{sato_sasaki}. 
The situation after the first set of investigations in the eighties has been nicely reviewed by den Nijs\cite{Den_Nijs}. Evidence in favour of a direct transition close to the Potts point, hence of a Lifshitz point different from Potts, has been obtained by several authors\cite{Selke1982,houlrik1986,Duxbury,bartelt}, with a Lifshitz point around $\Delta\simeq 0.4$, but the nature of this transition could not be established. The scaling of the correlation length looked isotropic, with an exponent $\nu$ consistent with Potts, but at the same time an exponent $\bar \beta \simeq 0.8\pm0.1$ was reported by Duxbury et al\cite{Duxbury}, suggesting that the condition $\bar \beta = \nu_x$ was satisfied, hence that the transition could be in the Huse-Fisher universality class. In parallel, evidence has been obtained on the quantum version of the model that, at the Lifshitz point, the scaling becomes anisotropic with $\nu_y=1$\cite{HOWES1983169,howes1983}. Considering the fact that the thermodynamic exponents were found to be the same as at the Potts point, Den Nijs concluded that the transition is direct but probably not chiral up to the Lifshitz point, and that the exponents change abruptly at the Lifshitz point. At that point, hyperscaling with $\nu_y=1$ and $\alpha=1/3$ leads to $\nu_x=2/3$, hence to anisotropic scaling.

Shortly after, a family of integrable models with complex Boltzmann weights has been discovered at finite chirality\cite{Baxter1989,albertini,mccoy}, with thermodynamic exponents again consistent with Potts, and with anisotropic scaling with correlation exponents $1$ and $2/3$ respectively. This result has been generalized to the self-dual version of the model by Cardy\cite{cardy}, who was the first to show that these peculiar exponents can be realized as soon as the chirality is switched on. However, the model studied by Cardy has complex asymmetry parameters in both directions, and, as acknowledged in his paper, the generalization of his renormalization group calculation to the more physical model of Eq. \ref{eq:model} is far from straightforward.

Finally, motivated by recent experiments on Rydberg atoms, quantum models in 1D have been studied with DMRG, and evidence of anisotropic scaling with a dynamical exponent that increases continuously from $z=1$ at the Potts point has been reported\cite{samajdar,sachdev_dual}.

\begin{figure}[t!]
\includegraphics[width=0.49\textwidth]{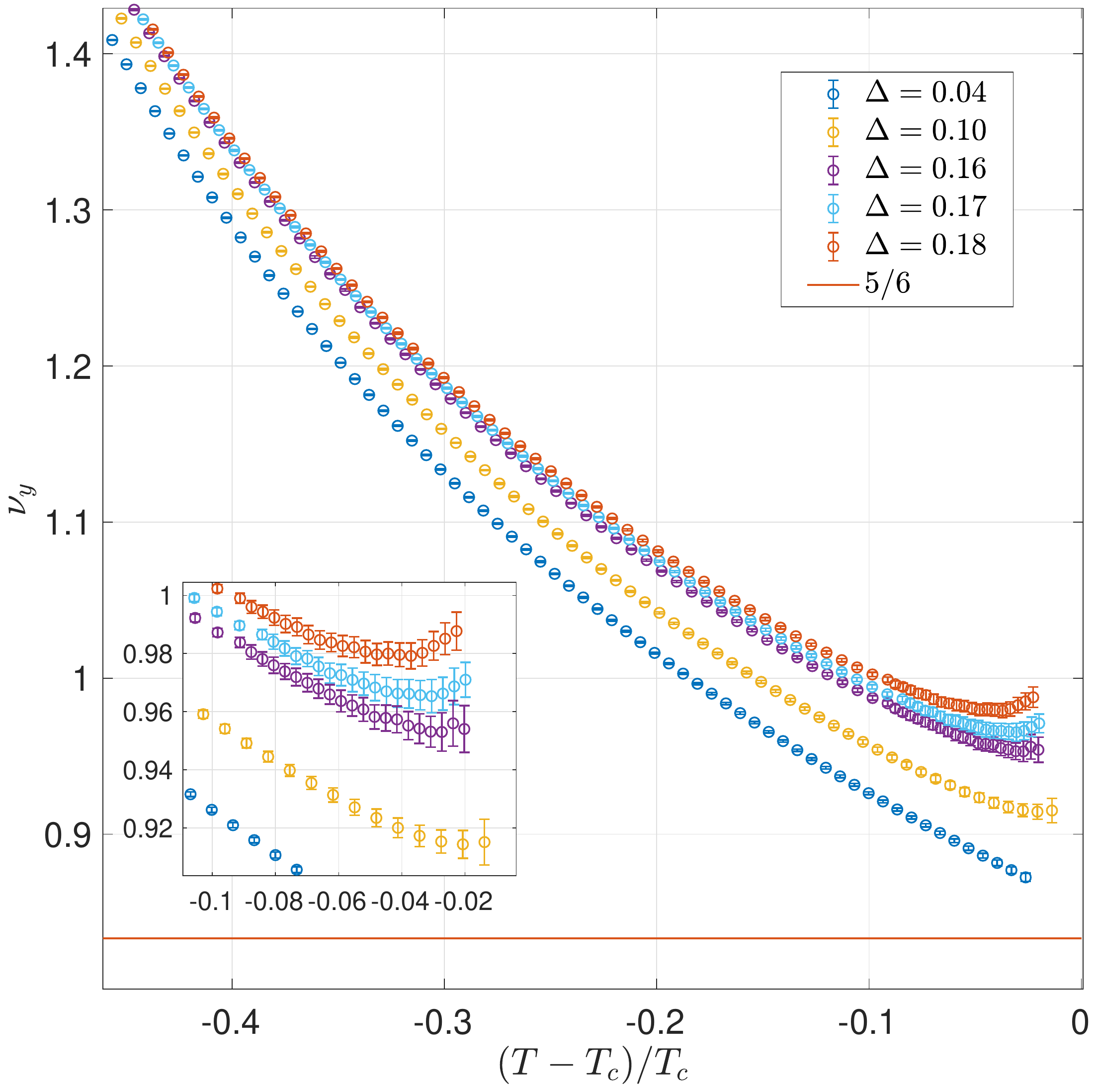}
\caption{Effective exponent $\nu_y$ for several values of $\Delta$, showing the development of a minimum followed by an upturn towards $T_c$. For small values of $\Delta$, the upturn cannot be observed, and the apparent limit of the exponent increases smoothly with $\Delta$.}
\label{fig:crossover}
\end{figure}

We believe that all these results can be understood in terms of the universality class we propose because of very strong crossover effects. 
Indeed, for the chiral Potts model, the crossover exponent is believed to be equal to 1/6\cite{nijs1984,stella1987}, i.e. very small. This implies that the critical regime where the true critical exponents can be observed scales as $(\Delta/\Delta_0)^6$, where $\Delta_0$ is a scaling factor, hence will be very narrow. This is clearly consistent with the results reported by Howes for a quantum version of the model\cite{howes1983}, who found that anisotropic scaling was only visible upon approaching the Lifshitz point. In our simulations, this is also what we observe. The scaling becomes clearly anisotropic around $\Delta=0.1$, but it is only close to the Lifshitz point that $\nu_y$ becomes consistent with 1. As can be seen in Fig.\ref{fig:crossover}, in which we show  the low temperature effective exponent $\nu_y$ for vertical cuts and for several values of $\Delta$, an upturn takes place at small $|t|$ when $\Delta$ is close enough to the Lifshitz point, and it is only this upturn that makes this exponent consistent with $\nu_y=1$. If we assume that upon approaching the Lifshitz point at $\Delta\approx 0.17$ the critical regime is of the order of $|t|\approx 0.03$, as suggested by the minimum of the effective exponent
$\nu_y$ (see Fig.\ref{fig:crossover}), we get $\Delta_0=0.3$. Then, for reduced temperatures above say $|t|\simeq0.01$, the upturn is not visible for $\Delta<0.15$, and the exponent apparently extrapolates to a value that increases progressively from the Potts value 5/6. In other words, corrections to scaling associated to this crossover exponent will be very large except very close to the Lifshitz point. 

By contrast, the exponent $\bar \beta$ does not seem to be affected by the proximity of the Potts point since its value is immediately very different, leading to what is maybe the most solid and most innovative result of the present paper, the evidence of an exponent $\bar \beta$ around 2/3 close to the Potts point, very far from the exact results $\bar \beta = 5/3$ at the Potts point and $\bar \beta = 1/2$ for the Pokrovsky-Talapov transition. Together with the evidence that $\nu_x$ also reaches a value consistent with $2/3$ for large enough $\Delta$, when the crossover can be overcome, this result points to the unique set of exponents we suggest all the way between the Potts point and the Lifshitz point.

Finally, we note that, at the Pokrosky-Talapov transition, the corrections to scaling appear to be quite small for $\nu_y$ and $\nu_x$. This is the basis of our determination of the Lifshitz point at $\Delta\simeq 0.17$. Note that above that value, and at least up to $\Delta=0.28$, the evidence we got in favor of an intermediate critical phase is only indirect because this phase is too narrow to be detected, as also suggested by previous investigations, which could not detect the intermediate critical phase below $\Delta\simeq 0.4$.

\subsection{Experimental implications}

For overlayers of adsorbates on surface, for which the model of Eq. (1) is of direct relevance, our results strongly suggest that, if the commensurate-incommensurate transition is a direct one between an ordered and a disordered phase, the critical exponents should be given by $\nu_y=1$, $\nu_x=2/3$, and $\bar \beta=2/3$. Experimental results obtained in the nineties on reconstructed surfaces in which the top layer plays the role of the overlayer are in very good agreement with these values\cite{abernathy,SelkeExperiment}. In particular, for the (3x1) reconstruction of the Si(113) surface, Abernathy et al \cite{abernathy} reported the values $\nu_y=1.06\pm0.07$, $\nu_x=0.65\pm 0.07$, and $\bar \beta=0.66\pm 0.05$, in excellent agreement with the present results. 

More recently, chains of Rydberg atoms have been shown to develop commensurate phases with periodicities given by $p=2, 3, 4,...$. For $p\ge 3$, the transition out of these phases is an example of a commensurate-incommensurate transition. So the universality class is expected to be the same as that of the chiral p-state Potts model. In the context of Rydberg chains\cite{lukin2017,lukin2019}, what can be conveniently measured is the Kibble-Zurek exponent $\mu$ that describes the growth of ordered domains as a function of the rate across a continuous phase transition. It is related to the dynamical exponent $z$ and to the correlation length exponent $\nu$ along the chain (the equivalent of $\nu_x$ in our case) by $\mu=\nu/(1+z \nu)$. Our results $z=3/2$ and $\nu=2/3$ lead to $\mu=1/3$, in reasonable agreement with the experimental results\cite{lukin2019}, which report values between 0.3 and 0.38.

\section{Summary}

We have revisited the phase diagram of the chiral three-state Potts model using a tensor network method introduced in the late nineties, the Corner Transfer Matrix Renormalization Group. The approximation on which it relies, a truncation of the size of the tensors during the contraction, is completely different from those used in the eighties to study the problem, statistical sampling or finite-size approaches. The specificity and main advantage of this method is that it naturally leads to extremely precise values of the correlation lengths and of the wave vector of the incommensurate phase in the accessible temperature range so that a systematic investigation of effective exponents and of their evolution upon approaching the critical temperature is possible. Combined with the prediction by Huse and Fisher that, away from the Potts point, the transition is either a direct transition in a new universality class or a Pokrovsky-Talapov transition into a critical phase followed by a Kosterlitz-Thouless transition, this approach has allowed us to give strong numerical evidence in favour of an intermediate region of chiral transition between the Potts point and a Lifshitz point at $\Delta\simeq 0.17$. In particular, close but away from the Potts point, there is compelling evidence that the incommensurability exponent $\bar \beta$ has a value around 2/3, very far from the exact results $\bar \beta = 5/3$ at the Potts point and $\bar \beta = 1/2$ for the Pokrovsky-Talapov transition. Moreover, the critical exponents that we have extracted from this analysis are consistent with those exactly known for the chiral transition of the self dual version of the model, $\nu_x=\bar \beta=2/3$, $\nu_y=1$, $\alpha=1/3$, and $z=3/2$, suggesting that the chiral transition is governed by the same universality class in all these models. These exponents are also fully consistent with experiments from the nineties on reconstructed surfaces, and in reasonable agreement with recent experiments on chains of Rydberg atoms. It would be interesting to see if this universality class can be established using field theory arguments. A first step in that direction has been taken recently by Whitsitt, Samajdar, and Sachdev\cite{sachdev_dual}, who performed an expansion around a dual version of the chiral 4-state Potts model in dimension 3 using $\delta=4-N$ and $\epsilon=3-d$ as expansion parameters to describe the chiral $N=3$-state Potts model in dimension $d=2$. Their results for the dynamical exponent $z$ up to next-to-leading order $z=1.57$ and for the correlation length exponent $\nu$ up to leading order $\nu=0.6$ are in reasonable agreement with our results.

{\it Acknowledgments.} We thank Natalia Chepiga, Paul Fendley, and Frank Verstraete for very useful discussions, and Titouan Dorthe, Matthjis Hogervorst, and Joao Penedones for a collaboration on a related topic.
This work has been supported by the Swiss National Science Foundation.
The calculations have been performed using the facilities of the Scientific IT and Application Support Center of EPFL.

\bibliography{bibliography}

\providecommand{\noopsort}[1]{}\providecommand{\singleletter}[1]{#1}%
\begin{thebibliography}{57}%
\makeatletter
\providecommand \@ifxundefined [1]{%
 \@ifx{#1\undefined}
}%
\providecommand \@ifnum [1]{%
 \ifnum #1\expandafter \@firstoftwo
 \else \expandafter \@secondoftwo
 \fi
}%
\providecommand \@ifx [1]{%
 \ifx #1\expandafter \@firstoftwo
 \else \expandafter \@secondoftwo
 \fi
}%
\providecommand \natexlab [1]{#1}%
\providecommand \enquote  [1]{``#1''}%
\providecommand \bibnamefont  [1]{#1}%
\providecommand \bibfnamefont [1]{#1}%
\providecommand \citenamefont [1]{#1}%
\providecommand \href@noop [0]{\@secondoftwo}%
\providecommand \href [0]{\begingroup \@sanitize@url \@href}%
\providecommand \@href[1]{\@@startlink{#1}\@@href}%
\providecommand \@@href[1]{\endgroup#1\@@endlink}%
\providecommand \@sanitize@url [0]{\catcode `\\12\catcode `\$12\catcode
  `\&12\catcode `\#12\catcode `\^12\catcode `\_12\catcode `\%12\relax}%
\providecommand \@@startlink[1]{}%
\providecommand \@@endlink[0]{}%
\providecommand \url  [0]{\begingroup\@sanitize@url \@url }%
\providecommand \@url [1]{\endgroup\@href {#1}{\urlprefix }}%
\providecommand \urlprefix  [0]{URL }%
\providecommand \Eprint [0]{\href }%
\providecommand \doibase [0]{https://doi.org/}%
\providecommand \selectlanguage [0]{\@gobble}%
\providecommand \bibinfo  [0]{\@secondoftwo}%
\providecommand \bibfield  [0]{\@secondoftwo}%
\providecommand \translation [1]{[#1]}%
\providecommand \BibitemOpen [0]{}%
\providecommand \bibitemStop [0]{}%
\providecommand \bibitemNoStop [0]{.\EOS\space}%
\providecommand \EOS [0]{\spacefactor3000\relax}%
\providecommand \BibitemShut  [1]{\csname bibitem#1\endcsname}%
\let\auto@bib@innerbib\@empty
\bibitem [{\citenamefont {Ostlund}(1981)}]{ostlund}%
  \BibitemOpen
  \bibfield  {author} {\bibinfo {author} {\bibfnamefont {S.}~\bibnamefont
  {Ostlund}},\ }\bibfield  {title} {\bibinfo {title} {Incommensurate and
  commensurate phases in asymmetric clock models},\ }\href
  {https://doi.org/10.1103/PhysRevB.24.398} {\bibfield  {journal} {\bibinfo
  {journal} {Phys. Rev. B}\ }\textbf {\bibinfo {volume} {24}},\ \bibinfo
  {pages} {398} (\bibinfo {year} {1981})}\BibitemShut {NoStop}%
\bibitem [{\citenamefont {Huse}(1981)}]{huse}%
  \BibitemOpen
  \bibfield  {author} {\bibinfo {author} {\bibfnamefont {D.~A.}\ \bibnamefont
  {Huse}},\ }\bibfield  {title} {\bibinfo {title} {Simple three-state model
  with infinitely many phases},\ }\href
  {https://doi.org/10.1103/PhysRevB.24.5180} {\bibfield  {journal} {\bibinfo
  {journal} {Phys. Rev. B}\ }\textbf {\bibinfo {volume} {24}},\ \bibinfo
  {pages} {5180} (\bibinfo {year} {1981})}\BibitemShut {NoStop}%
\bibitem [{\citenamefont {Huse}\ and\ \citenamefont
  {Fisher}(1982)}]{HuseFisher}%
  \BibitemOpen
  \bibfield  {author} {\bibinfo {author} {\bibfnamefont {D.~A.}\ \bibnamefont
  {Huse}}\ and\ \bibinfo {author} {\bibfnamefont {M.~E.}\ \bibnamefont
  {Fisher}},\ }\bibfield  {title} {\bibinfo {title} {Domain walls and the
  melting of commensurate surface phases},\ }\href
  {https://doi.org/10.1103/PhysRevLett.49.793} {\bibfield  {journal} {\bibinfo
  {journal} {Phys. Rev. Lett.}\ }\textbf {\bibinfo {volume} {49}},\ \bibinfo
  {pages} {793} (\bibinfo {year} {1982})}\BibitemShut {NoStop}%
\bibitem [{\citenamefont {Selke}\ and\ \citenamefont
  {Yeomans}(1982)}]{Selke1982}%
  \BibitemOpen
  \bibfield  {author} {\bibinfo {author} {\bibfnamefont {W.}~\bibnamefont
  {Selke}}\ and\ \bibinfo {author} {\bibfnamefont {J.~M.}\ \bibnamefont
  {Yeomans}},\ }\bibfield  {title} {\bibinfo {title} {A monte carlo study of
  the asymmetric clock or chiral potts model in two dimensions},\ }\href
  {https://doi.org/10.1007/BF01307706} {\bibfield  {journal} {\bibinfo
  {journal} {Zeitschrift f{\"u}r Physik B Condensed Matter}\ }\textbf {\bibinfo
  {volume} {46}},\ \bibinfo {pages} {311} (\bibinfo {year} {1982})}\BibitemShut
  {NoStop}%
\bibitem [{\citenamefont {Haldane}\ \emph {et~al.}(1983)\citenamefont
  {Haldane}, \citenamefont {Bak},\ and\ \citenamefont {Bohr}}]{haldane_bak}%
  \BibitemOpen
  \bibfield  {author} {\bibinfo {author} {\bibfnamefont {F.~D.~M.}\
  \bibnamefont {Haldane}}, \bibinfo {author} {\bibfnamefont {P.}~\bibnamefont
  {Bak}},\ and\ \bibinfo {author} {\bibfnamefont {T.}~\bibnamefont {Bohr}},\
  }\bibfield  {title} {\bibinfo {title} {Phase diagrams of surface structures
  from bethe-ansatz solutions of the quantum sine-gordon model},\ }\href
  {https://doi.org/10.1103/PhysRevB.28.2743} {\bibfield  {journal} {\bibinfo
  {journal} {Phys. Rev. B}\ }\textbf {\bibinfo {volume} {28}},\ \bibinfo
  {pages} {2743} (\bibinfo {year} {1983})}\BibitemShut {NoStop}%
\bibitem [{\citenamefont {Schulz}(1983)}]{schulz}%
  \BibitemOpen
  \bibfield  {author} {\bibinfo {author} {\bibfnamefont {H.~J.}\ \bibnamefont
  {Schulz}},\ }\bibfield  {title} {\bibinfo {title} {Phase transitions in
  monolayers adsorbed on uniaxial substrates},\ }\href
  {https://doi.org/10.1103/PhysRevB.28.2746} {\bibfield  {journal} {\bibinfo
  {journal} {Phys. Rev. B}\ }\textbf {\bibinfo {volume} {28}},\ \bibinfo
  {pages} {2746} (\bibinfo {year} {1983})}\BibitemShut {NoStop}%
\bibitem [{\citenamefont {Duxbury}\ \emph {et~al.}(1984)\citenamefont
  {Duxbury}, \citenamefont {Yeomans},\ and\ \citenamefont {Beale}}]{Duxbury}%
  \BibitemOpen
  \bibfield  {author} {\bibinfo {author} {\bibfnamefont {P.~M.}\ \bibnamefont
  {Duxbury}}, \bibinfo {author} {\bibfnamefont {J.}~\bibnamefont {Yeomans}},\
  and\ \bibinfo {author} {\bibfnamefont {P.~D.}\ \bibnamefont {Beale}},\
  }\bibfield  {title} {\bibinfo {title} {Wavevector scaling and the phase
  diagram of the chiral clock model},\ }\href
  {http://stacks.iop.org/0305-4470/17/i=4/a=005} {\bibfield  {journal}
  {\bibinfo  {journal} {Journal of Physics A: Mathematical and General}\
  }\textbf {\bibinfo {volume} {17}},\ \bibinfo {pages} {L179} (\bibinfo {year}
  {1984})}\BibitemShut {NoStop}%
\bibitem [{\citenamefont {den Nijs}(1984)}]{nijs1984}%
  \BibitemOpen
  \bibfield  {author} {\bibinfo {author} {\bibfnamefont {M.}~\bibnamefont {den
  Nijs}},\ }\bibfield  {title} {\bibinfo {title} {Extended scaling relations
  for the chiral and cubic crossover exponents},\ }\href
  {https://doi.org/10.1088/0305-4470/17/5/015} {\bibfield  {journal} {\bibinfo
  {journal} {Journal of Physics A: Mathematical and General}\ }\textbf
  {\bibinfo {volume} {17}},\ \bibinfo {pages} {L295} (\bibinfo {year}
  {1984})}\BibitemShut {NoStop}%
\bibitem [{\citenamefont {Huse}\ and\ \citenamefont
  {Fisher}(1984)}]{HuseFisher1984}%
  \BibitemOpen
  \bibfield  {author} {\bibinfo {author} {\bibfnamefont {D.~A.}\ \bibnamefont
  {Huse}}\ and\ \bibinfo {author} {\bibfnamefont {M.~E.}\ \bibnamefont
  {Fisher}},\ }\bibfield  {title} {\bibinfo {title} {Commensurate melting,
  domain walls, and dislocations},\ }\href
  {https://doi.org/10.1103/PhysRevB.29.239} {\bibfield  {journal} {\bibinfo
  {journal} {Phys. Rev. B}\ }\textbf {\bibinfo {volume} {29}},\ \bibinfo
  {pages} {239} (\bibinfo {year} {1984})}\BibitemShut {NoStop}%
\bibitem [{\citenamefont {Yeomans}\ and\ \citenamefont
  {Derrida}(1985)}]{yeomans1985}%
  \BibitemOpen
  \bibfield  {author} {\bibinfo {author} {\bibfnamefont {J.}~\bibnamefont
  {Yeomans}}\ and\ \bibinfo {author} {\bibfnamefont {B.}~\bibnamefont
  {Derrida}},\ }\bibfield  {title} {\bibinfo {title} {Bulk and interface
  scaling properties of the chiral clock model},\ }\href
  {https://doi.org/10.1088/0305-4470/18/12/031} {\bibfield  {journal} {\bibinfo
   {journal} {Journal of Physics A: Mathematical and General}\ }\textbf
  {\bibinfo {volume} {18}},\ \bibinfo {pages} {2343} (\bibinfo {year}
  {1985})}\BibitemShut {NoStop}%
\bibitem [{\citenamefont {Houlrik}\ and\ \citenamefont
  {Jensen}(1986)}]{houlrik1986}%
  \BibitemOpen
  \bibfield  {author} {\bibinfo {author} {\bibfnamefont {J.~M.}\ \bibnamefont
  {Houlrik}}\ and\ \bibinfo {author} {\bibfnamefont {S.~J.~K.}\ \bibnamefont
  {Jensen}},\ }\bibfield  {title} {\bibinfo {title} {Phase diagram of the
  three-state chiral clock model studied by monte carlo renormalization-group
  calculations},\ }\href {https://doi.org/10.1103/PhysRevB.34.325} {\bibfield
  {journal} {\bibinfo  {journal} {Phys. Rev. B}\ }\textbf {\bibinfo {volume}
  {34}},\ \bibinfo {pages} {325} (\bibinfo {year} {1986})}\BibitemShut
  {NoStop}%
\bibitem [{\citenamefont {Bartelt}\ \emph {et~al.}(1987)\citenamefont
  {Bartelt}, \citenamefont {Einstein},\ and\ \citenamefont
  {Roelofs}}]{bartelt}%
  \BibitemOpen
  \bibfield  {author} {\bibinfo {author} {\bibfnamefont {N.~C.}\ \bibnamefont
  {Bartelt}}, \bibinfo {author} {\bibfnamefont {T.~L.}\ \bibnamefont
  {Einstein}},\ and\ \bibinfo {author} {\bibfnamefont {L.~D.}\ \bibnamefont
  {Roelofs}},\ }\bibfield  {title} {\bibinfo {title} {Structure factors
  associated with the melting of a (31) ordered phase on a centered-rectangular
  lattice gas: Effective scaling in a three-state chiral-clock-like model},\
  }\href {https://doi.org/10.1103/PhysRevB.35.4812} {\bibfield  {journal}
  {\bibinfo  {journal} {Phys. Rev. B}\ }\textbf {\bibinfo {volume} {35}},\
  \bibinfo {pages} {4812} (\bibinfo {year} {1987})}\BibitemShut {NoStop}%
\bibitem [{\citenamefont {Stella}\ \emph {et~al.}(1987)\citenamefont {Stella},
  \citenamefont {Xie}, \citenamefont {Einstein},\ and\ \citenamefont
  {Bartelt}}]{stella1987}%
  \BibitemOpen
  \bibfield  {author} {\bibinfo {author} {\bibfnamefont {A.~L.}\ \bibnamefont
  {Stella}}, \bibinfo {author} {\bibfnamefont {X.~C.}\ \bibnamefont {Xie}},
  \bibinfo {author} {\bibfnamefont {T.~L.}\ \bibnamefont {Einstein}},\ and\
  \bibinfo {author} {\bibfnamefont {N.~C.}\ \bibnamefont {Bartelt}},\
  }\bibfield  {title} {\bibinfo {title} {Wavevector scaling, surface critical
  behavior, and wetting in the 2-d, 3-state chiral clock model},\ }\href
  {https://doi.org/10.1007/BF01307260} {\bibfield  {journal} {\bibinfo
  {journal} {Zeitschrift f{\"u}r Physik B Condensed Matter}\ }\textbf {\bibinfo
  {volume} {67}},\ \bibinfo {pages} {357} (\bibinfo {year} {1987})}\BibitemShut
  {NoStop}%
\bibitem [{\citenamefont {Au-Yang}\ \emph {et~al.}(1987)\citenamefont
  {Au-Yang}, \citenamefont {McCoy}, \citenamefont {Perk}, \citenamefont
  {Tang},\ and\ \citenamefont {Yan}}]{auyang1987}%
  \BibitemOpen
  \bibfield  {author} {\bibinfo {author} {\bibfnamefont {H.}~\bibnamefont
  {Au-Yang}}, \bibinfo {author} {\bibfnamefont {B.~M.}\ \bibnamefont {McCoy}},
  \bibinfo {author} {\bibfnamefont {J.~H.}\ \bibnamefont {Perk}}, \bibinfo
  {author} {\bibfnamefont {S.}~\bibnamefont {Tang}},\ and\ \bibinfo {author}
  {\bibfnamefont {M.-L.}\ \bibnamefont {Yan}},\ }\bibfield  {title} {\bibinfo
  {title} {Commuting transfer matrices in the chiral potts models: Solutions of
  star-triangle equations with genus>1},\ }\href
  {https://doi.org/https://doi.org/10.1016/0375-9601(87)90065-X} {\bibfield
  {journal} {\bibinfo  {journal} {Physics Letters A}\ }\textbf {\bibinfo
  {volume} {123}},\ \bibinfo {pages} {219 } (\bibinfo {year}
  {1987})}\BibitemShut {NoStop}%
\bibitem [{\citenamefont {Baxter}\ \emph {et~al.}(1988)\citenamefont {Baxter},
  \citenamefont {Perk},\ and\ \citenamefont {Au-Yang}}]{baxter1988}%
  \BibitemOpen
  \bibfield  {author} {\bibinfo {author} {\bibfnamefont {R.}~\bibnamefont
  {Baxter}}, \bibinfo {author} {\bibfnamefont {J.}~\bibnamefont {Perk}},\ and\
  \bibinfo {author} {\bibfnamefont {H.}~\bibnamefont {Au-Yang}},\ }\bibfield
  {title} {\bibinfo {title} {New solutions of the star-triangle relations for
  the chiral potts model},\ }\href
  {https://doi.org/https://doi.org/10.1016/0375-9601(88)90896-1} {\bibfield
  {journal} {\bibinfo  {journal} {Physics Letters A}\ }\textbf {\bibinfo
  {volume} {128}},\ \bibinfo {pages} {138 } (\bibinfo {year}
  {1988})}\BibitemShut {NoStop}%
\bibitem [{\citenamefont {Everts}\ and\ \citenamefont
  {Roder}(1989)}]{Everts_1989}%
  \BibitemOpen
  \bibfield  {author} {\bibinfo {author} {\bibfnamefont {H.~U.}\ \bibnamefont
  {Everts}}\ and\ \bibinfo {author} {\bibfnamefont {H.}~\bibnamefont {Roder}},\
  }\bibfield  {title} {\bibinfo {title} {Transfer matrix study of the chiral
  clock model in the hamiltonian limit},\ }\href
  {https://doi.org/10.1088/0305-4470/22/13/040} {\bibfield  {journal} {\bibinfo
   {journal} {Journal of Physics A: Mathematical and General}\ }\textbf
  {\bibinfo {volume} {22}},\ \bibinfo {pages} {2475} (\bibinfo {year}
  {1989})}\BibitemShut {NoStop}%
\bibitem [{\citenamefont {Baxter}(1989)}]{Baxter1989}%
  \BibitemOpen
  \bibfield  {author} {\bibinfo {author} {\bibfnamefont {R.~J.}\ \bibnamefont
  {Baxter}},\ }\bibfield  {title} {\bibinfo {title} {Superintegrable chiral
  potts model: Thermodynamic properties, an ``inverse'' model, and a simple
  associated hamiltonian},\ }\href {https://doi.org/10.1007/BF01023632}
  {\bibfield  {journal} {\bibinfo  {journal} {Journal of Statistical Physics}\
  }\textbf {\bibinfo {volume} {57}},\ \bibinfo {pages} {1} (\bibinfo {year}
  {1989})}\BibitemShut {NoStop}%
\bibitem [{\citenamefont {Albertini}\ \emph {et~al.}(1989)\citenamefont
  {Albertini}, \citenamefont {McCoy}, \citenamefont {Perk},\ and\ \citenamefont
  {Tang}}]{albertini}%
  \BibitemOpen
  \bibfield  {author} {\bibinfo {author} {\bibfnamefont {G.}~\bibnamefont
  {Albertini}}, \bibinfo {author} {\bibfnamefont {B.~M.}\ \bibnamefont
  {McCoy}}, \bibinfo {author} {\bibfnamefont {J.~H.}\ \bibnamefont {Perk}},\
  and\ \bibinfo {author} {\bibfnamefont {S.}~\bibnamefont {Tang}},\ }\bibfield
  {title} {\bibinfo {title} {Excitation spectrum and order parameter for the
  integrable n-state chiral potts model},\ }\href
  {https://doi.org/https://doi.org/10.1016/0550-3213(89)90415-X} {\bibfield
  {journal} {\bibinfo  {journal} {Nuclear Physics B}\ }\textbf {\bibinfo
  {volume} {314}},\ \bibinfo {pages} {741 } (\bibinfo {year}
  {1989})}\BibitemShut {NoStop}%
\bibitem [{\citenamefont {McCoy}\ and\ \citenamefont {shyr
  Roan}(1990)}]{mccoy}%
  \BibitemOpen
  \bibfield  {author} {\bibinfo {author} {\bibfnamefont {B.~M.}\ \bibnamefont
  {McCoy}}\ and\ \bibinfo {author} {\bibfnamefont {S.}~\bibnamefont {shyr
  Roan}},\ }\bibfield  {title} {\bibinfo {title} {Excitation spectrum and phase
  structure of the chiral potts model},\ }\href
  {https://doi.org/https://doi.org/10.1016/0375-9601(90)90230-L} {\bibfield
  {journal} {\bibinfo  {journal} {Physics Letters A}\ }\textbf {\bibinfo
  {volume} {150}},\ \bibinfo {pages} {347 } (\bibinfo {year}
  {1990})}\BibitemShut {NoStop}%
\bibitem [{\citenamefont {Cardy}(1993)}]{cardy}%
  \BibitemOpen
  \bibfield  {author} {\bibinfo {author} {\bibfnamefont {J.~L.}\ \bibnamefont
  {Cardy}},\ }\bibfield  {title} {\bibinfo {title} {Critical exponents of the
  chiral potts model from conformal field theory},\ }\href
  {https://doi.org/https://doi.org/10.1016/0550-3213(93)90353-Q} {\bibfield
  {journal} {\bibinfo  {journal} {Nuclear Physics B}\ }\textbf {\bibinfo
  {volume} {389}},\ \bibinfo {pages} {577 } (\bibinfo {year}
  {1993})}\BibitemShut {NoStop}%
\bibitem [{\citenamefont {Au-Yang}\ and\ \citenamefont
  {Perk}(1996)}]{auyang1996}%
  \BibitemOpen
  \bibfield  {author} {\bibinfo {author} {\bibfnamefont {H.}~\bibnamefont
  {Au-Yang}}\ and\ \bibinfo {author} {\bibfnamefont {J.~H.}\ \bibnamefont
  {Perk}},\ }\bibfield  {title} {\bibinfo {title} {Phase diagram in the
  generalized chiral clock models},\ }\href
  {https://doi.org/https://doi.org/10.1016/S0378-4371(96)00058-1} {\bibfield
  {journal} {\bibinfo  {journal} {Physica A: Statistical Mechanics and its
  Applications}\ }\textbf {\bibinfo {volume} {228}},\ \bibinfo {pages} {78 }
  (\bibinfo {year} {1996})}\BibitemShut {NoStop}%
\bibitem [{\citenamefont {Sato}\ and\ \citenamefont
  {Sasaki}(2000)}]{sato_sasaki}%
  \BibitemOpen
  \bibfield  {author} {\bibinfo {author} {\bibfnamefont {H.}~\bibnamefont
  {Sato}}\ and\ \bibinfo {author} {\bibfnamefont {K.}~\bibnamefont {Sasaki}},\
  }\bibfield  {title} {\bibinfo {title} {Numerical study of the two-dimensional
  three-state chiral clock model by the density matrix renormalization group
  method},\ }\href {https://doi.org/10.1143/JPSJ.69.1050} {\bibfield  {journal}
  {\bibinfo  {journal} {Journal of the Physical Society of Japan}\ }\textbf
  {\bibinfo {volume} {69}},\ \bibinfo {pages} {1050} (\bibinfo {year}
  {2000})}\BibitemShut {NoStop}%
\bibitem [{\citenamefont {Fendley}(2014)}]{fendley_parafermions}%
  \BibitemOpen
  \bibfield  {author} {\bibinfo {author} {\bibfnamefont {P.}~\bibnamefont
  {Fendley}},\ }\bibfield  {title} {\bibinfo {title} {Free parafermions},\
  }\href {http://stacks.iop.org/1751-8121/47/i=7/a=075001} {\bibfield
  {journal} {\bibinfo  {journal} {Journal of Physics A: Mathematical and
  Theoretical}\ }\textbf {\bibinfo {volume} {47}},\ \bibinfo {pages} {075001}
  (\bibinfo {year} {2014})}\BibitemShut {NoStop}%
\bibitem [{\citenamefont {Whitsitt}\ \emph {et~al.}(2018)\citenamefont
  {Whitsitt}, \citenamefont {Samajdar},\ and\ \citenamefont
  {Sachdev}}]{sachdev_dual}%
  \BibitemOpen
  \bibfield  {author} {\bibinfo {author} {\bibfnamefont {S.}~\bibnamefont
  {Whitsitt}}, \bibinfo {author} {\bibfnamefont {R.}~\bibnamefont {Samajdar}},\
  and\ \bibinfo {author} {\bibfnamefont {S.}~\bibnamefont {Sachdev}},\
  }\bibfield  {title} {\bibinfo {title} {Quantum field theory for the chiral
  clock transition in one spatial dimension},\ }\href
  {https://doi.org/10.1103/PhysRevB.98.205118} {\bibfield  {journal} {\bibinfo
  {journal} {Phys. Rev. B}\ }\textbf {\bibinfo {volume} {98}},\ \bibinfo
  {pages} {205118} (\bibinfo {year} {2018})}\BibitemShut {NoStop}%
\bibitem [{\citenamefont {Centen}\ \emph {et~al.}(1982)\citenamefont {Centen},
  \citenamefont {Rittenberg},\ and\ \citenamefont {Marcu}}]{CENTEN1982585}%
  \BibitemOpen
  \bibfield  {author} {\bibinfo {author} {\bibfnamefont {P.}~\bibnamefont
  {Centen}}, \bibinfo {author} {\bibfnamefont {V.}~\bibnamefont {Rittenberg}},\
  and\ \bibinfo {author} {\bibfnamefont {M.}~\bibnamefont {Marcu}},\ }\bibfield
   {title} {\bibinfo {title} {Non-universality in z3 symmetric spin systems},\
  }\href {https://doi.org/https://doi.org/10.1016/0550-3213(82)90079-7}
  {\bibfield  {journal} {\bibinfo  {journal} {Nuclear Physics B}\ }\textbf
  {\bibinfo {volume} {205}},\ \bibinfo {pages} {585 } (\bibinfo {year}
  {1982})}\BibitemShut {NoStop}%
\bibitem [{\citenamefont {Howes}\ \emph {et~al.}(1983)\citenamefont {Howes},
  \citenamefont {Kadanoff},\ and\ \citenamefont {Nijs}}]{HOWES1983169}%
  \BibitemOpen
  \bibfield  {author} {\bibinfo {author} {\bibfnamefont {S.}~\bibnamefont
  {Howes}}, \bibinfo {author} {\bibfnamefont {L.~P.}\ \bibnamefont
  {Kadanoff}},\ and\ \bibinfo {author} {\bibfnamefont {M.~D.}\ \bibnamefont
  {Nijs}},\ }\bibfield  {title} {\bibinfo {title} {Quantum model for
  commensurate-incommensurate transitions},\ }\href
  {https://doi.org/https://doi.org/10.1016/0550-3213(83)90212-2} {\bibfield
  {journal} {\bibinfo  {journal} {Nuclear Physics B}\ }\textbf {\bibinfo
  {volume} {215}},\ \bibinfo {pages} {169 } (\bibinfo {year}
  {1983})}\BibitemShut {NoStop}%
\bibitem [{\citenamefont {Howes}(1983)}]{howes1983}%
  \BibitemOpen
  \bibfield  {author} {\bibinfo {author} {\bibfnamefont {S.~F.}\ \bibnamefont
  {Howes}},\ }\bibfield  {title} {\bibinfo {title} {Commensurate-incommensurate
  transitions and the lifshitz point in the quantum asymmetric clock model},\
  }\href {https://doi.org/10.1103/PhysRevB.27.1762} {\bibfield  {journal}
  {\bibinfo  {journal} {Phys. Rev. B}\ }\textbf {\bibinfo {volume} {27}},\
  \bibinfo {pages} {1762} (\bibinfo {year} {1983})}\BibitemShut {NoStop}%
\bibitem [{\citenamefont {{von Gehlen}}\ and\ \citenamefont
  {Rittenberg}(1985)}]{vongehlen}%
  \BibitemOpen
  \bibfield  {author} {\bibinfo {author} {\bibfnamefont {G.}~\bibnamefont {{von
  Gehlen}}}\ and\ \bibinfo {author} {\bibfnamefont {V.}~\bibnamefont
  {Rittenberg}},\ }\bibfield  {title} {\bibinfo {title} {Zn-symmetric quantum
  chains with an infinite set of conserved charges and zn zero modes},\ }\href
  {https://doi.org/https://doi.org/10.1016/0550-3213(85)90350-5} {\bibfield
  {journal} {\bibinfo  {journal} {Nuclear Physics B}\ }\textbf {\bibinfo
  {volume} {257}},\ \bibinfo {pages} {351 } (\bibinfo {year}
  {1985})}\BibitemShut {NoStop}%
\bibitem [{\citenamefont {Fendley}\ \emph {et~al.}(2004)\citenamefont
  {Fendley}, \citenamefont {Sengupta},\ and\ \citenamefont
  {Sachdev}}]{fendley}%
  \BibitemOpen
  \bibfield  {author} {\bibinfo {author} {\bibfnamefont {P.}~\bibnamefont
  {Fendley}}, \bibinfo {author} {\bibfnamefont {K.}~\bibnamefont {Sengupta}},\
  and\ \bibinfo {author} {\bibfnamefont {S.}~\bibnamefont {Sachdev}},\
  }\bibfield  {title} {\bibinfo {title} {Competing density-wave orders in a
  one-dimensional hard-boson model},\ }\href
  {https://doi.org/10.1103/PhysRevB.69.075106} {\bibfield  {journal} {\bibinfo
  {journal} {Phys. Rev. B}\ }\textbf {\bibinfo {volume} {69}},\ \bibinfo
  {pages} {075106} (\bibinfo {year} {2004})}\BibitemShut {NoStop}%
\bibitem [{\citenamefont {Sachdev}\ \emph {et~al.}(2002)\citenamefont
  {Sachdev}, \citenamefont {Sengupta},\ and\ \citenamefont
  {Girvin}}]{sachdev_girvin}%
  \BibitemOpen
  \bibfield  {author} {\bibinfo {author} {\bibfnamefont {S.}~\bibnamefont
  {Sachdev}}, \bibinfo {author} {\bibfnamefont {K.}~\bibnamefont {Sengupta}},\
  and\ \bibinfo {author} {\bibfnamefont {S.~M.}\ \bibnamefont {Girvin}},\
  }\bibfield  {title} {\bibinfo {title} {Mott insulators in strong electric
  fields},\ }\href {https://doi.org/10.1103/PhysRevB.66.075128} {\bibfield
  {journal} {\bibinfo  {journal} {Phys. Rev. B}\ }\textbf {\bibinfo {volume}
  {66}},\ \bibinfo {pages} {075128} (\bibinfo {year} {2002})}\BibitemShut
  {NoStop}%
\bibitem [{\citenamefont {Lesanovsky}(2012)}]{Lesanovsky2012}%
  \BibitemOpen
  \bibfield  {author} {\bibinfo {author} {\bibfnamefont {I.}~\bibnamefont
  {Lesanovsky}},\ }\bibfield  {title} {\bibinfo {title} {Liquid ground state,
  gap, and excited states of a strongly correlated spin chain},\ }\href
  {https://doi.org/10.1103/PhysRevLett.108.105301} {\bibfield  {journal}
  {\bibinfo  {journal} {Phys. Rev. Lett.}\ }\textbf {\bibinfo {volume} {108}},\
  \bibinfo {pages} {105301} (\bibinfo {year} {2012})}\BibitemShut {NoStop}%
\bibitem [{\citenamefont {Zhuang}\ \emph {et~al.}(2015)\citenamefont {Zhuang},
  \citenamefont {Changlani}, \citenamefont {Tubman},\ and\ \citenamefont
  {Hughes}}]{hughes}%
  \BibitemOpen
  \bibfield  {author} {\bibinfo {author} {\bibfnamefont {Y.}~\bibnamefont
  {Zhuang}}, \bibinfo {author} {\bibfnamefont {H.~J.}\ \bibnamefont
  {Changlani}}, \bibinfo {author} {\bibfnamefont {N.~M.}\ \bibnamefont
  {Tubman}},\ and\ \bibinfo {author} {\bibfnamefont {T.~L.}\ \bibnamefont
  {Hughes}},\ }\bibfield  {title} {\bibinfo {title} {Phase diagram of the
  ${Z}_{3}$ parafermionic chain with chiral interactions},\ }\href
  {https://doi.org/10.1103/PhysRevB.92.035154} {\bibfield  {journal} {\bibinfo
  {journal} {Phys. Rev. B}\ }\textbf {\bibinfo {volume} {92}},\ \bibinfo
  {pages} {035154} (\bibinfo {year} {2015})}\BibitemShut {NoStop}%
\bibitem [{\citenamefont {Bernien}\ \emph {et~al.}(2017)\citenamefont
  {Bernien}, \citenamefont {Schwartz}, \citenamefont {Keesling}, \citenamefont
  {Levine}, \citenamefont {Omran}, \citenamefont {Pichler}, \citenamefont
  {Choi}, \citenamefont {Zibrov}, \citenamefont {Endres}, \citenamefont
  {Greiner}, \citenamefont {Vuletic},\ and\ \citenamefont {Lukin}}]{lukin2017}%
  \BibitemOpen
  \bibfield  {author} {\bibinfo {author} {\bibfnamefont {H.}~\bibnamefont
  {Bernien}}, \bibinfo {author} {\bibfnamefont {S.}~\bibnamefont {Schwartz}},
  \bibinfo {author} {\bibfnamefont {A.}~\bibnamefont {Keesling}}, \bibinfo
  {author} {\bibfnamefont {H.}~\bibnamefont {Levine}}, \bibinfo {author}
  {\bibfnamefont {A.}~\bibnamefont {Omran}}, \bibinfo {author} {\bibfnamefont
  {H.}~\bibnamefont {Pichler}}, \bibinfo {author} {\bibfnamefont
  {S.}~\bibnamefont {Choi}}, \bibinfo {author} {\bibfnamefont {A.~S.}\
  \bibnamefont {Zibrov}}, \bibinfo {author} {\bibfnamefont {M.}~\bibnamefont
  {Endres}}, \bibinfo {author} {\bibfnamefont {M.}~\bibnamefont {Greiner}},
  \bibinfo {author} {\bibfnamefont {V.}~\bibnamefont {Vuletic}},\ and\ \bibinfo
  {author} {\bibfnamefont {M.~D.}\ \bibnamefont {Lukin}},\ }\bibfield  {title}
  {\bibinfo {title} {Probing many-body dynamics on a 51-atom quantum
  simulator},\ }\href {http://dx.doi.org/10.1038/nature24622} {\bibfield
  {journal} {\bibinfo  {journal} {Nature}\ }\textbf {\bibinfo {volume} {551}},\
  \bibinfo {pages} {579} (\bibinfo {year} {2017})}\BibitemShut {NoStop}%
\bibitem [{\citenamefont {Samajdar}\ \emph {et~al.}(2018)\citenamefont
  {Samajdar}, \citenamefont {Choi}, \citenamefont {Pichler}, \citenamefont
  {Lukin},\ and\ \citenamefont {Sachdev}}]{samajdar}%
  \BibitemOpen
  \bibfield  {author} {\bibinfo {author} {\bibfnamefont {R.}~\bibnamefont
  {Samajdar}}, \bibinfo {author} {\bibfnamefont {S.}~\bibnamefont {Choi}},
  \bibinfo {author} {\bibfnamefont {H.}~\bibnamefont {Pichler}}, \bibinfo
  {author} {\bibfnamefont {M.~D.}\ \bibnamefont {Lukin}},\ and\ \bibinfo
  {author} {\bibfnamefont {S.}~\bibnamefont {Sachdev}},\ }\bibfield  {title}
  {\bibinfo {title} {Numerical study of the chiral ${\mathbb{z}}_{3}$ quantum
  phase transition in one spatial dimension},\ }\href
  {https://doi.org/10.1103/PhysRevA.98.023614} {\bibfield  {journal} {\bibinfo
  {journal} {Phys. Rev. A}\ }\textbf {\bibinfo {volume} {98}},\ \bibinfo
  {pages} {023614} (\bibinfo {year} {2018})}\BibitemShut {NoStop}%
\bibitem [{\citenamefont {{Keesling}}\ \emph {et~al.}(2019)\citenamefont
  {{Keesling}}, \citenamefont {{Omran}}, \citenamefont {{Levine}},
  \citenamefont {{Bernien}}, \citenamefont {{Pichler}}, \citenamefont {{Choi}},
  \citenamefont {{Samajdar}}, \citenamefont {{Schwartz}}, \citenamefont
  {{Silvi}}, \citenamefont {{Sachdev}}, \citenamefont {{Zoller}}, \citenamefont
  {{Endres}}, \citenamefont {{Greiner}}, \citenamefont {{Vuleti{\'c}}},
  \citenamefont {{}},\ and\ \citenamefont {{Lukin}}}]{lukin2019}%
  \BibitemOpen
  \bibfield  {author} {\bibinfo {author} {\bibfnamefont {A.}~\bibnamefont
  {{Keesling}}}, \bibinfo {author} {\bibfnamefont {A.}~\bibnamefont {{Omran}}},
  \bibinfo {author} {\bibfnamefont {H.}~\bibnamefont {{Levine}}}, \bibinfo
  {author} {\bibfnamefont {H.}~\bibnamefont {{Bernien}}}, \bibinfo {author}
  {\bibfnamefont {H.}~\bibnamefont {{Pichler}}}, \bibinfo {author}
  {\bibfnamefont {S.}~\bibnamefont {{Choi}}}, \bibinfo {author} {\bibfnamefont
  {R.}~\bibnamefont {{Samajdar}}}, \bibinfo {author} {\bibfnamefont
  {S.}~\bibnamefont {{Schwartz}}}, \bibinfo {author} {\bibfnamefont
  {P.}~\bibnamefont {{Silvi}}}, \bibinfo {author} {\bibfnamefont
  {S.}~\bibnamefont {{Sachdev}}}, \bibinfo {author} {\bibfnamefont
  {P.}~\bibnamefont {{Zoller}}}, \bibinfo {author} {\bibfnamefont
  {M.}~\bibnamefont {{Endres}}}, \bibinfo {author} {\bibfnamefont
  {M.}~\bibnamefont {{Greiner}}}, \bibinfo {author} {\bibnamefont
  {{Vuleti{\'c}}}}, \bibinfo {author} {\bibfnamefont {V.}~\bibnamefont {{}}},\
  and\ \bibinfo {author} {\bibfnamefont {M.~D.}\ \bibnamefont {{Lukin}}},\
  }\bibfield  {title} {\bibinfo {title} {{Quantum Kibble-Zurek mechanism and
  critical dynamics on a programmable Rydberg simulator}},\ }\href
  {https://doi.org/10.1038/s41586-019-1070-1} {\bibfield  {journal} {\bibinfo
  {journal} {Nature}\ }\textbf {\bibinfo {volume} {568}},\ \bibinfo {pages}
  {207} (\bibinfo {year} {2019})}\BibitemShut {NoStop}%
\bibitem [{\citenamefont {Chepiga}\ and\ \citenamefont
  {Mila}(2019)}]{chepiga_mila_PRL}%
  \BibitemOpen
  \bibfield  {author} {\bibinfo {author} {\bibfnamefont {N.}~\bibnamefont
  {Chepiga}}\ and\ \bibinfo {author} {\bibfnamefont {F.}~\bibnamefont {Mila}},\
  }\bibfield  {title} {\bibinfo {title} {Floating phase versus chiral
  transition in a 1d hard-boson model},\ }\href
  {https://doi.org/10.1103/PhysRevLett.122.017205} {\bibfield  {journal}
  {\bibinfo  {journal} {Phys. Rev. Lett.}\ }\textbf {\bibinfo {volume} {122}},\
  \bibinfo {pages} {017205} (\bibinfo {year} {2019})}\BibitemShut {NoStop}%
\bibitem [{\citenamefont {Giudici}\ \emph {et~al.}(2019)\citenamefont
  {Giudici}, \citenamefont {Angelone}, \citenamefont {Magnifico}, \citenamefont
  {Zeng}, \citenamefont {Giudice}, \citenamefont {Mendes-Santos},\ and\
  \citenamefont {Dalmonte}}]{dalmonte2019}%
  \BibitemOpen
  \bibfield  {author} {\bibinfo {author} {\bibfnamefont {G.}~\bibnamefont
  {Giudici}}, \bibinfo {author} {\bibfnamefont {A.}~\bibnamefont {Angelone}},
  \bibinfo {author} {\bibfnamefont {G.}~\bibnamefont {Magnifico}}, \bibinfo
  {author} {\bibfnamefont {Z.}~\bibnamefont {Zeng}}, \bibinfo {author}
  {\bibfnamefont {G.}~\bibnamefont {Giudice}}, \bibinfo {author} {\bibfnamefont
  {T.}~\bibnamefont {Mendes-Santos}},\ and\ \bibinfo {author} {\bibfnamefont
  {M.}~\bibnamefont {Dalmonte}},\ }\bibfield  {title} {\bibinfo {title}
  {Diagnosing potts criticality and two-stage melting in one-dimensional
  hard-core boson models},\ }\href {https://doi.org/10.1103/PhysRevB.99.094434}
  {\bibfield  {journal} {\bibinfo  {journal} {Phys. Rev. B}\ }\textbf {\bibinfo
  {volume} {99}},\ \bibinfo {pages} {094434} (\bibinfo {year}
  {2019})}\BibitemShut {NoStop}%
\bibitem [{\citenamefont {Kosterlitz}\ and\ \citenamefont
  {Thouless}(1973)}]{Kosterlitz_Thouless}%
  \BibitemOpen
  \bibfield  {author} {\bibinfo {author} {\bibfnamefont {J.~M.}\ \bibnamefont
  {Kosterlitz}}\ and\ \bibinfo {author} {\bibfnamefont {D.~J.}\ \bibnamefont
  {Thouless}},\ }\bibfield  {title} {\bibinfo {title} {Ordering, metastability
  and phase transitions in two-dimensional systems},\ }\href
  {http://stacks.iop.org/0022-3719/6/i=7/a=010} {\bibfield  {journal} {\bibinfo
   {journal} {Journal of Physics C: Solid State Physics}\ }\textbf {\bibinfo
  {volume} {6}},\ \bibinfo {pages} {1181} (\bibinfo {year} {1973})}\BibitemShut
  {NoStop}%
\bibitem [{\citenamefont {Pokrovsky}\ and\ \citenamefont
  {Talapov}(1979)}]{Pokrovsky_Talapov}%
  \BibitemOpen
  \bibfield  {author} {\bibinfo {author} {\bibfnamefont {V.~L.}\ \bibnamefont
  {Pokrovsky}}\ and\ \bibinfo {author} {\bibfnamefont {A.~L.}\ \bibnamefont
  {Talapov}},\ }\bibfield  {title} {\bibinfo {title} {Ground state, spectrum,
  and phase diagram of two-dimensional incommensurate crystals},\ }\href
  {https://doi.org/10.1103/PhysRevLett.42.65} {\bibfield  {journal} {\bibinfo
  {journal} {Phys. Rev. Lett.}\ }\textbf {\bibinfo {volume} {42}},\ \bibinfo
  {pages} {65} (\bibinfo {year} {1979})}\BibitemShut {NoStop}%
\bibitem [{\citenamefont {Schulz}(1980)}]{schulz1980}%
  \BibitemOpen
  \bibfield  {author} {\bibinfo {author} {\bibfnamefont {H.~J.}\ \bibnamefont
  {Schulz}},\ }\bibfield  {title} {\bibinfo {title} {Critical behavior of
  commensurate-incommensurate phase transitions in two dimensions},\ }\href
  {https://doi.org/10.1103/PhysRevB.22.5274} {\bibfield  {journal} {\bibinfo
  {journal} {Phys. Rev. B}\ }\textbf {\bibinfo {volume} {22}},\ \bibinfo
  {pages} {5274} (\bibinfo {year} {1980})}\BibitemShut {NoStop}%
\bibitem [{\citenamefont {Pleimling}\ and\ \citenamefont
  {Selke}(1998)}]{pleimling_selke}%
  \BibitemOpen
  \bibfield  {author} {\bibinfo {author} {\bibfnamefont {M.}~\bibnamefont
  {Pleimling}}\ and\ \bibinfo {author} {\bibfnamefont {W.}~\bibnamefont
  {Selke}},\ }\bibfield  {title} {\bibinfo {title} {Critical phenomena at
  perfect and non-perfect surfaces},\ }\href
  {https://doi.org/10.1007/s100510050198} {\bibfield  {journal} {\bibinfo
  {journal} {The European Physical Journal B - Condensed Matter and Complex
  Systems}\ }\textbf {\bibinfo {volume} {1}},\ \bibinfo {pages} {385} (\bibinfo
  {year} {1998})}\BibitemShut {NoStop}%
\bibitem [{\citenamefont {Pleimling}\ and\ \citenamefont
  {Henkel}(2001)}]{pleimling_henkel}%
  \BibitemOpen
  \bibfield  {author} {\bibinfo {author} {\bibfnamefont {M.}~\bibnamefont
  {Pleimling}}\ and\ \bibinfo {author} {\bibfnamefont {M.}~\bibnamefont
  {Henkel}},\ }\bibfield  {title} {\bibinfo {title} {Anisotropic scaling and
  generalized conformal invariance at lifshitz points},\ }\href
  {https://doi.org/10.1103/PhysRevLett.87.125702} {\bibfield  {journal}
  {\bibinfo  {journal} {Phys. Rev. Lett.}\ }\textbf {\bibinfo {volume} {87}},\
  \bibinfo {pages} {125702} (\bibinfo {year} {2001})}\BibitemShut {NoStop}%
\bibitem [{\citenamefont {Nishino}\ and\ \citenamefont
  {Okunishi}(1996)}]{nishino}%
  \BibitemOpen
  \bibfield  {author} {\bibinfo {author} {\bibfnamefont {T.}~\bibnamefont
  {Nishino}}\ and\ \bibinfo {author} {\bibfnamefont {K.}~\bibnamefont
  {Okunishi}},\ }\bibfield  {title} {\bibinfo {title} {Corner transfer matrix
  renormalization group method},\ }\href {https://doi.org/10.1143/JPSJ.65.891}
  {\bibfield  {journal} {\bibinfo  {journal} {J. Phys. Soc. Jpn.}\ }\textbf
  {\bibinfo {volume} {65}},\ \bibinfo {pages} {891} (\bibinfo {year}
  {1996})}\BibitemShut {NoStop}%
\bibitem [{\citenamefont {Baxter}(1968)}]{baxter1968}%
  \BibitemOpen
  \bibfield  {author} {\bibinfo {author} {\bibfnamefont {R.~J.}\ \bibnamefont
  {Baxter}},\ }\bibfield  {title} {\bibinfo {title} {Dimers on a rectangular
  lattice},\ }\href {https://doi.org/10.1063/1.1664623} {\bibfield  {journal}
  {\bibinfo  {journal} {Journal of Mathematical Physics}\ }\textbf {\bibinfo
  {volume} {9}},\ \bibinfo {pages} {650} (\bibinfo {year} {1968})}\BibitemShut
  {NoStop}%
\bibitem [{\citenamefont {Baxter}(1978)}]{baxter1978}%
  \BibitemOpen
  \bibfield  {author} {\bibinfo {author} {\bibfnamefont {R.~J.}\ \bibnamefont
  {Baxter}},\ }\bibfield  {title} {\bibinfo {title} {Variational approximations
  for square lattice models in statistical mechanics},\ }\href
  {https://doi.org/10.1007/BF01011693} {\bibfield  {journal} {\bibinfo
  {journal} {Journal of Statistical Physics}\ }\textbf {\bibinfo {volume}
  {19}},\ \bibinfo {pages} {461} (\bibinfo {year} {1978})}\BibitemShut
  {NoStop}%
\bibitem [{\citenamefont {White}(1992)}]{dmrg1}%
  \BibitemOpen
  \bibfield  {author} {\bibinfo {author} {\bibfnamefont {S.~R.}\ \bibnamefont
  {White}},\ }\bibfield  {title} {\bibinfo {title} {Density matrix formulation
  for quantum renormalization groups},\ }\href
  {https://doi.org/10.1103/PhysRevLett.69.2863} {\bibfield  {journal} {\bibinfo
   {journal} {Phys. Rev. Lett.}\ }\textbf {\bibinfo {volume} {69}},\ \bibinfo
  {pages} {2863} (\bibinfo {year} {1992})}\BibitemShut {NoStop}%
\bibitem [{\citenamefont {White}(1993)}]{white1993}%
  \BibitemOpen
  \bibfield  {author} {\bibinfo {author} {\bibfnamefont {S.~R.}\ \bibnamefont
  {White}},\ }\bibfield  {title} {\bibinfo {title} {Density-matrix algorithms
  for quantum renormalization groups},\ }\href
  {https://doi.org/10.1103/PhysRevB.48.10345} {\bibfield  {journal} {\bibinfo
  {journal} {Phys. Rev. B}\ }\textbf {\bibinfo {volume} {48}},\ \bibinfo
  {pages} {10345} (\bibinfo {year} {1993})}\BibitemShut {NoStop}%
\bibitem [{\citenamefont {Or\'us}\ and\ \citenamefont {Vidal}(2009)}]{orus}%
  \BibitemOpen
  \bibfield  {author} {\bibinfo {author} {\bibfnamefont {R.}~\bibnamefont
  {Or\'us}}\ and\ \bibinfo {author} {\bibfnamefont {G.}~\bibnamefont {Vidal}},\
  }\bibfield  {title} {\bibinfo {title} {Simulation of two-dimensional quantum
  systems on an infinite lattice revisited: Corner transfer matrix for tensor
  contraction},\ }\href {https://doi.org/10.1103/PhysRevB.80.094403} {\bibfield
   {journal} {\bibinfo  {journal} {Phys. Rev. B}\ }\textbf {\bibinfo {volume}
  {80}},\ \bibinfo {pages} {094403} (\bibinfo {year} {2009})}\BibitemShut
  {NoStop}%
\bibitem [{\citenamefont {Rams}\ \emph {et~al.}(2018)\citenamefont {Rams},
  \citenamefont {Czarnik},\ and\ \citenamefont {Cincio}}]{czarnik2018}%
  \BibitemOpen
  \bibfield  {author} {\bibinfo {author} {\bibfnamefont {M.~M.}\ \bibnamefont
  {Rams}}, \bibinfo {author} {\bibfnamefont {P.}~\bibnamefont {Czarnik}},\ and\
  \bibinfo {author} {\bibfnamefont {L.}~\bibnamefont {Cincio}},\ }\bibfield
  {title} {\bibinfo {title} {Precise extrapolation of the correlation function
  asymptotics in uniform tensor network states with application to the
  bose-hubbard and xxz models},\ }\href
  {https://doi.org/10.1103/PhysRevX.8.041033} {\bibfield  {journal} {\bibinfo
  {journal} {Phys. Rev. X}\ }\textbf {\bibinfo {volume} {8}},\ \bibinfo {pages}
  {041033} (\bibinfo {year} {2018})}\BibitemShut {NoStop}%
\bibitem [{\citenamefont {Baxter}(1980)}]{baxter1980}%
  \BibitemOpen
  \bibfield  {author} {\bibinfo {author} {\bibfnamefont {R.~J.}\ \bibnamefont
  {Baxter}},\ }\bibfield  {title} {\bibinfo {title} {Hard hexagons: exact
  solution},\ }\href {https://doi.org/10.1088/0305-4470/13/3/007} {\bibfield
  {journal} {\bibinfo  {journal} {Journal of Physics A: Mathematical and
  General}\ }\textbf {\bibinfo {volume} {13}},\ \bibinfo {pages} {L61}
  (\bibinfo {year} {1980})}\BibitemShut {NoStop}%
\bibitem [{\citenamefont {Baxter}\ and\ \citenamefont
  {Pearce}(1982)}]{Baxter_Pearce}%
  \BibitemOpen
  \bibfield  {author} {\bibinfo {author} {\bibfnamefont {R.~J.}\ \bibnamefont
  {Baxter}}\ and\ \bibinfo {author} {\bibfnamefont {P.~A.}\ \bibnamefont
  {Pearce}},\ }\bibfield  {title} {\bibinfo {title} {Hard hexagons: interfacial
  tension and correlation length},\ }\href
  {http://stacks.iop.org/0305-4470/15/i=3/a=027} {\bibfield  {journal}
  {\bibinfo  {journal} {Journal of Physics A: Mathematical and General}\
  }\textbf {\bibinfo {volume} {15}},\ \bibinfo {pages} {897} (\bibinfo {year}
  {1982})}\BibitemShut {NoStop}%
\bibitem [{\citenamefont {Salas}\ and\ \citenamefont
  {Sokal}(1997)}]{sokal1997}%
  \BibitemOpen
  \bibfield  {author} {\bibinfo {author} {\bibfnamefont {J.}~\bibnamefont
  {Salas}}\ and\ \bibinfo {author} {\bibfnamefont {A.~D.}\ \bibnamefont
  {Sokal}},\ }\bibfield  {title} {\bibinfo {title} {Dynamic critical behavior
  of the swendsen-wang algorithm: The two-dimensional three-state potts model
  revisited},\ }\href {https://doi.org/10.1007/BF02181478} {\bibfield
  {journal} {\bibinfo  {journal} {Journal of Statistical Physics}\ }\textbf
  {\bibinfo {volume} {87}},\ \bibinfo {pages} {1} (\bibinfo {year}
  {1997})}\BibitemShut {NoStop}%
\bibitem [{Note1()}]{Note1}%
  \BibitemOpen
  \bibinfo {note} {Indeed, assuming that $e-e_c\propto t^{1-\alpha
  }(1+at^\theta )$ leads to $C\propto t^{-\alpha }(1+a(1+\theta /(1-\alpha
  ))t^\theta )$ so that the coefficient of the correction to scaling is
  increased by a factor $1+\theta /(1-\alpha )$.}\BibitemShut {Stop}%
\bibitem [{\citenamefont {Nienhuis}(1982)}]{nienhuis1982}%
  \BibitemOpen
  \bibfield  {author} {\bibinfo {author} {\bibfnamefont {B.}~\bibnamefont
  {Nienhuis}},\ }\bibfield  {title} {\bibinfo {title} {Analytical calculation
  of two leading exponents of the dilute potts model},\ }\href
  {https://doi.org/10.1088/0305-4470/15/1/028} {\bibfield  {journal} {\bibinfo
  {journal} {Journal of Physics A: Mathematical and General}\ }\textbf
  {\bibinfo {volume} {15}},\ \bibinfo {pages} {199} (\bibinfo {year}
  {1982})}\BibitemShut {NoStop}%
\bibitem [{\citenamefont {den Nijs}(1988)}]{Den_Nijs}%
  \BibitemOpen
  \bibfield  {author} {\bibinfo {author} {\bibfnamefont {M.}~\bibnamefont {den
  Nijs}},\ }\bibfield  {title} {\bibinfo {title} {The domain wall theory of
  two-dimensional commensurate-incommensurate phase transitions},\ }\href@noop
  {} {\bibfield  {journal} {\bibinfo  {journal} {Phase Transitions and Critical
  Phenomena}\ }\textbf {\bibinfo {volume} {12}},\ \bibinfo {pages} {219}
  (\bibinfo {year} {1988})}\BibitemShut {NoStop}%
\bibitem [{\citenamefont {Abernathy}\ \emph {et~al.}(1994)\citenamefont
  {Abernathy}, \citenamefont {Song}, \citenamefont {Blum}, \citenamefont
  {Birgeneau},\ and\ \citenamefont {Mochrie}}]{abernathy}%
  \BibitemOpen
  \bibfield  {author} {\bibinfo {author} {\bibfnamefont {D.~L.}\ \bibnamefont
  {Abernathy}}, \bibinfo {author} {\bibfnamefont {S.}~\bibnamefont {Song}},
  \bibinfo {author} {\bibfnamefont {K.~I.}\ \bibnamefont {Blum}}, \bibinfo
  {author} {\bibfnamefont {R.~J.}\ \bibnamefont {Birgeneau}},\ and\ \bibinfo
  {author} {\bibfnamefont {S.~G.~J.}\ \bibnamefont {Mochrie}},\ }\bibfield
  {title} {\bibinfo {title} {Chiral melting of the si(113) (3$\times$1)
  reconstruction},\ }\href {https://doi.org/10.1103/PhysRevB.49.2691}
  {\bibfield  {journal} {\bibinfo  {journal} {Phys. Rev. B}\ }\textbf {\bibinfo
  {volume} {49}},\ \bibinfo {pages} {2691} (\bibinfo {year}
  {1994})}\BibitemShut {NoStop}%
\bibitem [{\citenamefont {Schreiner}\ \emph {et~al.}(1994)\citenamefont
  {Schreiner}, \citenamefont {Jacobi},\ and\ \citenamefont
  {Selke}}]{SelkeExperiment}%
  \BibitemOpen
  \bibfield  {author} {\bibinfo {author} {\bibfnamefont {J.}~\bibnamefont
  {Schreiner}}, \bibinfo {author} {\bibfnamefont {K.}~\bibnamefont {Jacobi}},\
  and\ \bibinfo {author} {\bibfnamefont {W.}~\bibnamefont {Selke}},\ }\bibfield
   {title} {\bibinfo {title} {Experimental evidence for chiral melting of the
  ge(113) and si(113) 3$\times$1 surface phases},\ }\href
  {https://doi.org/10.1103/PhysRevB.49.2706} {\bibfield  {journal} {\bibinfo
  {journal} {Phys. Rev. B}\ }\textbf {\bibinfo {volume} {49}},\ \bibinfo
  {pages} {2706} (\bibinfo {year} {1994})}\BibitemShut {NoStop}%
\end{thebibliography}%

\end{document}